\documentclass[a4paper,11pt,final]{article}
\usepackage[left=2.8cm,right=2.8cm,top=2.8cm,bottom=2.8cm]{geometry}
\usepackage{graphicx}
\usepackage{dcolumn}
\usepackage{bm}

\usepackage[utf8]{inputenc}
\usepackage[english]{babel}
\usepackage[T1]{fontenc}
\usepackage{mathptmx}

\usepackage{amsmath, amsfonts, amsthm, amssymb, amsbsy}
\usepackage{bm}
\usepackage{accents,comment}

\usepackage{xargs}
\usepackage[pdftex,dvipsnames]{xcolor}
\usepackage[colorinlistoftodos,prependcaption,obeyFinal]{todonotes}

\usepackage[pdftex,unicode]{hyperref}
\hypersetup{breaklinks=true}

\newcommand{\cc}{\mathbf{c}}
\newcommand{\dd}{\mathbf{d}}
\newcommand{\mm}{\mathbf{m}}

\newcommand{\rr}{\mathbf{r}}
\newcommand{\pp}{\mathbf{p}}
\newcommand{\qq}{\mathbf{q}}

\newcommand{\vv}{\mathbf{v}}
\newcommand{\xx}{\mathbf{x}}
\newcommand{\yy}{\mathbf{y}}

\newcommand{\BB}{\mathbf{B}}

\newcommand{\II}{\mathbf{I}}
\newcommand{\LL}{\mathbf{L}}
\newcommand{\MM}{\mathbf{M}}

\newcommand{\RR}{\mathbf{R}}
\newcommand{\XX}{\mathbf{X}}

\renewcommand{\d}{\mathrm{d}}

\newcommand{\AAA}{\mathbb{A}}
\newcommand{\BBB}{\mathbb{B}}

\newcommand{\DDD}{\mathbb{D}}
\newcommand{\FFF}{\mathbb{F}}
\newcommand{\GGG}{\mathbb{G}}
\newcommand{\III}{\mathbb{I}}
\newcommand{\LLL}{\mathbb{L}}

\newcommand{\NNN}{\mathbb{N}}

\newcommand{\RRR}{\mathbb{R}}

\newcommand{\TTT}{\mathbb{T}}

\newcommand{\CCCC}{\mathcal{C}}
\newcommand{\NNNN}{\mathcal{N}}
\newcommand{\RRRR}{\mathcal{R}}

\newcommand{\Tirr}{\TTT_{\text{irr}}}
\newcommand{\Tel}{\TTT_{\text{el}}}
\newcommand{\el}{\text{el}}
\newcommand{\rev}{\text{rev}}
\newcommand{\irr}{\text{irr}}

\newcommand{\pd}[2]{\frac{\partial #1}{\partial #2}}
\newcommand{\pder}[2]{\pd{#1}{#2}}
\newcommand{\fd}[2]{\frac{\delta #1}{\delta #2}}
\newcommand{\fder}[2]{\fd{#1}{#2}}

\newcommand{\tens}[1]{\mathbb{#1}}

\newcommand{\F}{\tens{F}}
\newcommand{\I}{\tens{I}}
\renewcommand{\G}{\tens{G}}

\newcommand{\B}{\tens{B}}
\renewcommand{\L}{\tens{L}}
\newcommand{\D}{\tens{D}}

\newcommand{\kpt}{_{\kappa}}

\newcommand{\Fp}{\F\kpt}
\newcommand{\Lp}{\L\kpt}
\newcommand{\Dp}{\D\kpt}
\newcommand{\Bp}{\B\kpt}

\newcommand{\diag}{\mathrm{diag}}

\newcommand{\fibR}{\mathbf{N}}
\newcommand{\fibRi}{N}
\newcommand{\fibN}{{\mathbf{N}\kpt}}

\newcommand{\dotfibN}{{\dot{\mathbf{N}}\kpt}}

\newcommand{\fibC}{\mathbf{n}}
\newcommand{\fibCi}{n}

\newcommand{\nablasym}{\nabla_{\text{sym}}}

\DeclareMathOperator{\tr}{Tr}

\DeclareMathOperator{\dive}{div}

\newtheorem{remark}{Remark}

\begin{document}

\title{On compatibility of the Natural configuration framework with GENERIC:\\ Derivation of anisotropic rate-type models}

\author{P. Pelech,
K. Tůma,
M. Šípka,
M. Sýkora,
M. Pavelka
}

\date{\today}

\maketitle

\begin{abstract}
Within the framework of natural configurations developed by Rajagopal and Srinivasa, evolution within continuum thermodynamics is formulated as evolution of a natural configuration linked with the current configuration. 
On the other hand, withing the General Equation for Non-Equilibrium Reversible-Irreversible Coupling (GENERIC) framework,
the evolution is split into Hamiltonian mechanics and (generalized) gradient dynamics. 
These seemingly radically different approaches have actually a lot in common and we show their compatibility on a wide range of models.
Both frameworks are illustrated on~isotropic and anisotropic rate-type fluid models.
We propose an interpretation of the natural configurations within GENERIC and vice versa (when possible).
\end{abstract}


\section*{Introduction}
Non-equilibrium continuum thermodynamics
is a~li\-ve\-ly evolving scientific field with various
schools, frameworks, and theories that are partly compatible and partly in disagreement.
Mentioning just some of the possible approaches, the Classical Irreversible Thermodynamics \cite{dgm},
Rational Thermodynamics \cite{Truesdell-RT},
Extended Irreversible Thermodynamics \cite{Jou-EIT},
Rational Extended Thermodynamics \cite{Muller-Ruggeri,Muller-Thermodynamics},
Steepest Entropy Ascent \cite{SEA},
Internal Variables Theory \cite{Van-Berezovski},
Principle of Virtual Power \cite{Germain1998,Fremond2002,GurtinAnand2010,IsolaCorteGiorgio2017},
Symmetric Hyperbolic Thermodynamically Compatible (SHTC) equations \cite{GodR,GodRom2003,Peshkov-unified,ADER},
GENERIC and metriplectic systems \cite{HCO,PavelkaKlikaGrmela2018,BE, Morrison-brackets},
the framework of Rajagopal and Srinivasa (NCF) and entropy production maximization;
\cite{RajagopalSrinivasa2000,malek.j.prusa.v:derivation}.
Our goal is to compare the latter two approaches in detail. Although they might seem as rather different at first sight, we show that they are compatible in many cases.

Inspired by \cite{Leonov},
Rajagopal and Srinivasa used the same equations as in \cite{WapperomHulsen1998},
but brought up a new understanding of the equations and unknowns;
paper \cite{RajagopalSrinivasa2000} was chosen as the best paper in 2000 in the journal.
A~partial generalization of NCF for a~subclass of~anisotropic visco-elastic fluid followed shortly in \cite{RajagopalSrinivasa2001}.
The~main idea of this framework is that the overall evolution is split into the dissipative (irreversible) evolution from a reference to a natural configuration and elastic (reversible) evolution from the natural to the current configuration.

Such split resembles the way evolution equations are generated within the GENERIC framework,
where the reversible part is given by Hamiltonian mechanics and the irreversible by (generalized) gradient dynamics.
It was shown in \cite{besseling,Hutter2008a,Hutter2008b} that visco-elasto-plastic solids can indeed be formulated within GENERIC,
and a comprehensive review has been given in \cite{Hutter2009}.
In that review one can also find a statistical derivation of the evolution equations for the deformation tensor
while fluctuations of the deformation tensor where studied in \cite{Hutter-distortion}.
In \cite{Grmela-extensions} the evolution of the field of labels and other fields coming from the three-particle kinetic theory were derived.
Anisotropy in the dissipative part was described by means of statistical mechanics in \cite{Hutter2015}. GENERIC evolution of a director field was formulated in \cite{BE}, p. 528, but (when dropping the momentum of rotation) it only consists of advection of each of the three components of the directors as if they were scalar quantities. A geometric way towards advection of vector and covector fields, using the theory of semidirect products, was shown in \cite{VagnerPavelkaEsen2021}.

Our intention is to go beyond those works
in~the~following sense:
    (i) item using the distortion (inverse of deformation gradient)
		rather than the deformation tensor itself or the field of labels,
    (ii) by explicitly considering general evolution of transversally anisotropic systems (having the orientation vector as an extra state variable), and
    (iii) and by providing a new interpretation of the GENERIC evolution for visco-elasto-plastic materials based on the concept of natural configuration.

As a common ground for modeling within the two frameworks, we choose isotropic and anisotropic visco-elasto-plastic materials
(one may think e.g. about polymeric fluids, liquid crystals, or crystal plasticity).
For the sake of simplicity, we stay restricted 
to a very simple case of anisotropy -- the transverse isotropy.
In both frameworks, we derive new kinematics and present two specific models,
where in NCF we extend the model derived by \cite{RajagopalSrinivasa2001}.

Novelty of this manuscript lies in the following points:
(i) Both frameworks are compared in detail and interpreted within each other.
(ii) The anisotropic model within the NCF is refined.
(iii) A new derivation of the Poisson bracket for the anisotropic fluid is shown (including the distortion).
(iv) A formulation of irreversible dynamics of anisotropic media is presented based on dissipation potentials and Cauchy stress.

The structure of the article is the following.
In Section \ref{sec:toy} we explain the fundamental principles of GENERIC on~a~one particle example,
as this framework is supposed to be less known than other approaches based on~balance laws.
In Section \ref{sec:iso} we introduce both frameworks in~a~continuum setting
and compare them on the well known variant of the isotropic Giesekus model. 
Then in Section \ref{sec:aniso} we formulate anisotropic models in both frameworks.

\section{One-Particle Dynamics in GENERIC} \label{sec:toy}
The GENERIC framework will be illustrated on an as simple as possible toy model -- damped harmonic oscillator (e.g. a weight on a spring experiencing friction).
We start by explaining how the evolution equations are derived,
and how this procedure can be simplified.
We close the part by showing a modification suitable for a description
of isothermal processes.

The~first step we have to make is to~define the~\emph{state variables}, denoted here by $\qq$,
which describe the oscillator appropriately.
The state variables characterize the \emph{level of description} (the manifold of state variables).
Here we choose the~position of~the~weight
$\rr \in \RRR^3$,
its momentum $\pp \in (\RRR^3)^*$(dual space to $\RRR^3$)
and its total energy $E \in \RRR$.
This is the so called \emph{entropic representation},
where the fundamental thermodynamic relation for entropy
$S = S(\qq) = S(\rr,\pp,E)$ completely specifies
the 'material properties' of the weight and the spring;
see e.g. \cite{Callen}.
On the other hand, one may use the \emph{energetic representation},
where the state variables are $\rr$, $\pp$ and $S$.
The fundamental thermodynamic relation is then $E = E(\qq) = E(\rr,\pp,S)$
and can be obtained by inverting the entropic one. 

Note, however, that it is not always possible to switch between the representations and derive energy from entropy or vice versa.
In kinetic theory, where the state variable is the one-particle distribution function $f(\rr,\pp)$,
the energy is simply the kinetic energy while entropy is the Boltzmann entropy and they can not be obtained from each other.
This is also a difference between
the framework of Beris and Edwards \cite{BE},
where only one functional is needed to generate the dynamics,
and GENERIC where both entropy $S$ energy $E$ are needed.
In the present paper, however, we always have energy or entropy as a state variable and so we can switch between the representations.

Once having the state variables, energy, and entropy, we can proceed to formulate the evolution equations of the state variables.
The key idea is to compose the evolution equations from the \emph{reversible evolution} (representing mechanics)
and \emph{irreversible evolution} (representing thermodynamics).
For the former, we exploit the well developed machinery of Hamiltonian mechanics
while the latter is written as gradient dynamics;
see e.g. \cite{PavelkaKlikaGrmela2018} for more details.\footnote
{
	Especially when focused on differential geometry,
	the idea of having both Hamiltonian and thermodynamic evolution combined leads to the \emph{metriplectic systems} \cite{Morrison-brackets}.
	Such systems are equivalent to having GENERIC with a dissipative bracket instead of the dissipation potential.
}
In our setting we have
\begin{align*}
    \dot{\rr}
    &=
    \left( \dot{\rr} \right)_{\text{rev}}
    +
    \left( \dot{\rr} \right)_{\text{irr}}, \\
    \dot{\pp}
    &=
    \left( \dot{\pp} \right)_{\text{rev}}
    +
    \left( \dot{\pp} \right)_{\text{irr}}, \\
    \dot{S}
    &=
    \left( \dot{S} \right)_{\text{rev}}
    +
    \left( \dot{S} \right)_{\text{irr}},
\end{align*}
where the parts of the right hand side will be specified below.
One possibility of how to~distinguish between the reversible and irreversible part
is to observe the behavior with respect to the time reversal transformation (TRT),
which inverts the velocities of all particles; see \cite{PRE15}.

\subsection{Reversible Evolution}
Let us first recall Hamiltonian mechanics.
In general, a Poisson bracket is  
an~antisymmetric bilinear form taking functions (or functionals)
of the state variables as its arguments fulfilling also the Leibniz rule and Jacobi identity
\begin{subequations}
\begin{align*}
\{B,C\}=-\{C,B\} &\qquad \mbox{antisymmetry}\\
\{B,C+D\}=\{B,C\}+ \{B,D\}&\qquad \mbox{bilinearity}\\
\{B,C D\}=\{B,C\}D+C\{B,D\} &\qquad \mbox{Leibniz rule}\\
\{\{B,C\},D\}+
\{\{C,D\},B\}+
\{\{D,B\},C\}=0 &\qquad\mbox{Jacobi identity},
\end{align*}
\end{subequations}
see e.g. \cite{PavelkaKlikaGrmela2018}.

Consider now a~general set of state variables $\qq$.
Hamiltonian evolution of an arbitrary functional of the state variables $B(\qq)$ is then prescribed as 
\begin{equation*}
	(\dot{B})_{\text{rev}} = \{B,E\}, 
\end{equation*}
where $E$ stands for the energy (or Hamiltonian).
In particular, when localized, evolution of the state variables reads
\begin{equation} \label{eq:HamEvolV}
	(\dot{\qq})_{\text{rev}} = \{\qq,E\}.
\end{equation}
Both~these equations are in~a~coordinateless form.
For~explicit computations in~coordinates we introduce \emph{Poisson bivector}
$\LL := L^{ij} \frac{\partial}{\partial q^i}\otimes \frac{\partial}{\partial q^j}$,
a~twice contravariant antisymmetric tensor field whose coordinates are given by
\begin{equation*}
	L^{ij} := \{q^i, q^j\},
\end{equation*}
where $q^i$ and $q^j$ ranges over all state variables.
The~Poisson bracket is then equivalently described by Poisson bivector
\begin{equation*}
	\{B,C\} = \frac{\partial B}{\partial q^i} L^{ij} \frac{\partial C}{\partial q^j},
\end{equation*}
and the evolution equation \eqref{eq:HamEvolV} becomes
\begin{equation*}
	(\dot{q}^i)_{\text{rev}} = \{q^i, E\}=L^{ij}\frac{\partial E}{\partial q^j}.
\end{equation*}
Note that the derivative is to be understood as the functional derivative in general, particular realization of which is the partial derivative.

The antisymmetry of the Poisson bracket means that the energy is conserved automatically.
Assuming that we can multiply the state variables, the Leibniz rule makes the Hamiltonian evolution consistent with the Leibniz rule for the time derivative.
Jacobi identity can be interpreted as a self-consistency of the Hamiltonian evolution \cite{HCO}.
Namely, the right hand side of the equation can be seen as a component of the Hamiltonian vector field $\mathbf{X}_{E} = \mathbf{L}\cdot \d E$
($dE$ being the gradient of~energy),
and the~Jacobi identity is equivalent to~the Lie derivative of the Poisson bivector $\mathbf{L}$
with respect to the Hamiltonian vector field $\mathbf{X}_E$ being zero;
see e.g. \cite{Fecko}.
In other words, the Poisson bivector is advected by the Hamiltonian evolution,
which is the aforementioned self-consistency of Hamiltonian mechanics.

Let us now return to the simple example of the damped oscillator.
We shall adopt the energetic representation, i.e. the state variables $\rr, \pp$ and $S$, where $S$ is the entropy of the oscillator
(capturing the~possible heating up). 
Kinematics of the state variables is expressed by~\emph{canonical Poisson bracket} $\{\cdot,\cdot\}$
\begin{align} \label{eq:PB-can}
    \{B,C\}^{\text{can}}
    =
    \frac{\partial B}{\partial \rr} \cdot \frac{\partial C}{\partial \pp}
    -
    \frac{\partial C}{\partial \rr} \cdot \frac{\partial B}{\partial \pp},
    \qquad
    \forall B(\rr,\pp,S),\, C(\rr,\pp,S).
\end{align}
The Poisson bivector is just a~block-wise matrix in~$\RRR^{7 \times 7}$
\begin{equation*}
	\mathbf{L}
	=
	\begin{pmatrix}
        0 & \II & 0\\
        -\II & 0 & 0\\
        0 & 0 & 0
	\end{pmatrix},
\end{equation*}
where $\II$ stands for~identity on~$\RRR^3$.
Hamiltonian evolution of an arbitrary functional $B(\rr,\pp,S)$ is then given by
\begin{equation} \label{eq:HamCanEvo}
    \left( \dot{B} \right)_{\text{rev}}
    =
    \{B, E\}^{\text{can}}
    =
    \frac{\partial B}{\partial \rr} \cdot \frac{\partial E}{\partial \pp}
    -
    \frac{\partial E}{\partial \rr} \cdot \frac{\partial B}{\partial \pp},
\end{equation}
where $E(\rr,\pp,S)$ is the total energy of the system (or Hamiltonian).
The evolution equations for the state variables read
\begin{subequations}
\begin{align*}
   \left( \dot{\rr} \right)_{\text{rev}} &= \frac{\partial E}{\partial \pp},\nonumber\\
   \left( \dot{\pp} \right)_{\text{rev}} &= -\frac{\partial E}{\partial \rr},\nonumber\\
   \left( \dot{S} \right)_{\text{rev}} &= 0.\nonumber
\end{align*}
These are the usual Hamilton canonical equations equipped with the trivial evolution of entropy.
\end{subequations}

Since the~bracket \eqref{eq:PB-can} does not contain
partial derivatives with respect to~entropy,
entropy does not evolve in this setting at all.
This is a general principle within GENERIC: the reversible evolution does not change the entropy.
In other words, the entropy is always assumed to be a so called \emph{Casimir}
of the Poisson bracket $\{\cdot,\cdot\}$, i.e.
\begin{equation}\label{eq:deg-S}
    \forall E=E(\qq): \, \{S, E\} = 0,
    \quad
    \text{or in coordinates}
    \quad
    \forall j: \,
    \pder{S}{q^i} L^{ij} = 0.
\end{equation}
Entropy is thus a quantity intimately related to the geometry. 
At the same time, as already mentioned, the antisymmetry of the bracket implies
\begin{align*}
    \dot{E} = \{E, E\} = 0,
\end{align*}
i.e. the energy is automatically conserved.
This represents the \emph{first law of thermodynamics}. 

Making this example more explicit,
one may consider the usual form of energy
\begin{align} \label{eq:can-E}
    E(\qq) = E(\rr,\pp,S) = \frac{|\pp|^2}{2m}+V(\rr)+\epsilon(S),
\end{align}
which consists of kinetic ($m$ denotes the mass of the weight),
potential (typically $V(\rr) = \frac{1}{2}k|\rr|^2$, $k$ being the~``spring constant'')
and internal contributions.
The Hamiltonian evolution equations then gain the explicit form
\begin{subequations}
\begin{align*}
   (\dot{\rr})_{rev} &= \frac{\pp}{m}, \nonumber \\
   (\dot{\pp})_{rev} &= -\frac{\partial V}{\partial \rr}, \nonumber \\
   (\dot{S})_{rev} &= 0. \nonumber
\end{align*}
\end{subequations}

Let us briefly summarize the reversible kinematics.
Reversible evolution in GENERIC is generated by Hamiltonian mechanics, i.e. a Poisson bracket and energy.
The Poisson bracket corresponds to the choice of state variables and can be altered only by changing the variables.
The specific characters of the system (e.g. the material properties)
are specified by the energy as a function of the state variables.
Once it is
fixed, the reversible evolution equations can be written down in a closed form.
Moreover, entropy is required to be a~Casimir of the Poisson bracket,
and this degeneracy condition implies that the entropy of an isolated system is not altered by the Hamiltonian mechanics.
Finally, the energy of a closed system is automatically conserved due to the antisymmetry of the bracket.

\subsection{Irreversible Evolution}\label{subsec:toy-irr}
Let us now turn to the irreversible
 kinematics and dynamics.
Consider again an arbitrary set of state variables $\qq$.
Within GENERIC the irreversible evolution of the state variables is prescribed as a~(generalized) gradient dynamics,
see \cite{Gyarmati,Landau-Ginzburg}, i.e. as being generated by~a~dissipation potential $\Xi$,
\begin{equation}\label{eq.GD}
(\dot{\qq})_{\text{irr}} = \frac{\partial \Xi(\qq,\qq^*)}{\partial \qq^*}\Big|_{\qq^*= \frac{\partial S}{\partial \qq}}.
\end{equation}
The dissipation potential $\Xi$
depends on the state variables and on the entropic conjugate variables $\qq^*$, and it should satisfy the following criteria: 
(i) Positivity and with minimum at origin, 
          $\Xi \geq 0$ and $\frac{\partial \Xi}{\partial \qq^*}\Big|_{\qq^* =0}=0$.
(ii) Convexity (although this assumption can be weakened as shown in \cite{nonconvex}). 
(iii) Degeneracy so that mass and energy are conserved.
(iv) Even parity with respect to TRT.
The first criterion implies that the irreversible evolution vanishes
in the thermodynamic equilibrium $\qq^*= 0$.
The second combined with the first implies entropy is being produced.
In particular, the entropy of an isolated system grows, 
\begin{equation*}
	\dot{S}=\frac{\partial S}{\partial q^i}(\dot q^i)_{\text{irr}} = 
	\frac{\partial S}{\partial q^i}\frac{\partial \Xi(\qq,\qq^*)}{\partial \qq^*}\Big|_{\qq^*= \frac{\partial S}{\partial \qq}} \geq 0,
\end{equation*}
see e.g. \cite{JSP2020} for more details. The second law of thermodynamics is thus satisfied.
The fourth property assures that the evolution equations generated
by the dissipation potential are irreversible with respect to time-reversal transformation;
this makes the separation into the reversible and irreversible part unambiguous, see \cite{PRE15}.

In the original works on GENERIC \cite{Grmela1997,Ottinger1997} and in book \cite{HCO},
the irreversible evolution is expressed also by a dissipative bracket, see also \cite{Hutter2013},
which is equivalent to having a metriplectic system mentioned before.
This is recovered when the dissipation potential is quadratic, 
\begin{equation} \label{eq:DissPotQuad}
  \Xi(\qq,\qq^*) = \frac{1}{2} q^*_i M^{ij} q^*_j,
\end{equation}
where $M^{ij}$ is called the dissipation matrix; see \cite{Miroslav-guide}.
The dissipative evolution then becomes 
\begin{equation*}
	(\dot{q}^i)_{\text{irr}} = M^{ij}\frac{\partial S}{\partial q^j},
\end{equation*}
which is the gradient dynamics in the sense of \cite{Otto}.
Moreover, the dissipation matrix is positive definite (due to the convexity of $\Xi$),
and the degeneracy ensuring the conservation of energy reads
\begin{equation}\label{eq:deg-E}
	\frac{\partial E}{\partial q^i}M^{ij} = 0 \qquad\forall j.
\end{equation}
There is also a~tight connection between the dissipation matrix and fluctuations via the fluctuation-dissipation theorem;
see \cite{HCO}.\footnote
{
	The dissipation matrix is clearly symmetric when coming from a~dissipation potential,
	but this assumption was later relaxed in~\cite{HCO} to~allow also for non-symmetric matrices;
	however, there is an ongoing controversy about
	whether to allow for the non-symmetric dissipation operators,
	see e.g. the~discussion in~\cite{Hutter2013,Miroslav-WhyGeneric,PRE15,Mielke-Peletier,Mielke-Potential}.
}

Let us now turn to the simple example of~a~damped oscillator.
It is advantageous to choose the entropic representation, i.e. $\qq = (\rr,\pp,E)$,
since the gradient dynamics then attains a simpler form.
By inverting the fundamental thermodynamic relation \eqref{eq:can-E} we obtain
\begin{align} \label{eq:FTR}
    S = S(\qq) = S(\rr,\pp,E) = \sigma\left(E-\frac{|\pp|^2}{2m} - V(\rr)\right), 
\end{align}
which is equivalent to the original relation,
$\sigma
(\epsilon)
$
being the inverse function to the internal energy density $\epsilon(S)$. 
A~simple choice of~a~dissipation potential
corresponding to~friction is
\begin{equation*}
	\Xi(\qq,\qq^*) = \Xi(\rr,\pp,E,\rr^*,\pp^*,E^*) = \frac{1}{2} \zeta (
	\pp^*
	)^2.
\end{equation*}
The irreversible evolution is then 
\begin{subequations}
\begin{align}
    \label{eq:can-irr-a}
    \left( \dot{\rr} \right)_{\text{irr}}
    &=0\\
    \left( \dot{\pp} \right)_{\text{irr}}
    &=
    \left.
      \frac{\partial \Xi}{\partial \pp^*}
    \right|_{\pp^* = \frac{\partial S}{\partial \pp}}
	=\zeta \frac{\partial S}{\partial \pp} = -\frac{\zeta}{T}\frac{\pp}{m}, \\
    \label{eq:can-irr-c}
    \left( \dot{E} \right)_{\text{irr}}
    &=0,
\end{align}
\end{subequations}
where the derivative of entropy are eventually substituted for the conjugate variables.
With our choice of the dissipation potential $\Xi$ 
we have obtained a friction force,
which is proportional to the velocity
$\vv:=\pder{E}{\pp} = \pp/m$
via the friction coefficient $\zeta/T$, where
\begin{equation*}
    \frac{1}{T}
    :=
    \frac{\partial \sigma}{\partial\, \epsilon}
    =
    \frac{\partial S}{\partial E}
    =
    E^*.
\end{equation*}
is the inverse temperature.
The implied equation for entropy is
then
\begin{equation*}
(\dot{S})_{\text{irr}} = \frac{\partial S}{\partial \pp} \frac{\partial \Xi}{\partial \pp^*}
    \Big|_{\pp^* = \frac{\partial S}{\partial \pp}} = 
\frac{\zeta}{T^2}\left(\frac{\pp}{m}\right)^2\geq 0,
\end{equation*}
i.e.
 the second law
of thermodynamics
 is clearly satisfied. 

Since we already have the irreversible dynamics of $\rr$, $\pp$ and $S$,
we can readily add them to the reversible dynamics of the variables.
Note, however, that the gradient dynamics can be also transformed to the energetic representation easily;
see \cite{JSP2020,mielke-generic}.

\subsection{Final Equations}
Let us now combine the reversible and irreversible dynamics to the GENERIC set of evolution equations (again for a general set of state variables $\qq$), 
\begin{equation}\label{eq.GENERIC}
	\dot q^i = L^{ij} \frac{\partial E}{\partial q^j} + \frac{\partial \Xi}{\partial q^*_i}\Big|_{\qq^*= \frac{\partial S}{\partial \qq}}.
\end{equation}
The energy of an isolated system is conserved while its entropy is raised.
Using the dissipation matrix instead of a~general dissipation potential leads to
evolution equations
\begin{equation}\label{eq:GENERIC-mat}
\dot q^i = L^{ij} \frac{\partial E}{\partial q^j} + M^{ij}\frac{\partial S}{\partial q^j}.
\end{equation}

For our toy example we obtain (in the energetic representation)
\begin{subequations}
\begin{align*}
   \dot{\rr} &= \frac{\pp}{m},\nonumber\\
   \dot{\pp} &= -\frac{\partial V}{\partial \rr}-T^{-1} \zeta \frac{\pp}{m},\nonumber\\
   \dot{S} &= T^{-2}\zeta \left(\frac{\pp}{m}\right)^2.
\end{align*}
\end{subequations}
Evolution of the state variables is the sum of the reversible Hamiltonian evolution and irreversible gradient dynamics.
The Hamiltonian evolution conserves both energy and entropy while gradient dynamics conserves energy and produces entropy. It can be shown then the Onsager-Casimir reciprocal relations are satisfied automatically in a generalized sense \cite{HCO,PRE15,PavelkaKlikaGrmela2018}.

\subsection{Time-Reversal Transformation}
Let us now briefly return to the time-reversal transformation (TRT). This transformation inverts velocities of all particles, $\rr\rightarrow \rr$, $\pp\rightarrow -\pp$, $S\rightarrow S$, $E\rightarrow E$. Momentum is called odd with respect to TRT while the other variables even. If we apply TRT on the evolution equations, the time increment is also inverted and we obtain
\begin{subequations}
\begin{align*}
   -\dot{\rr} &= -\frac{\pp}{m},\nonumber\\
   \dot{\pp} &= -\frac{\partial V}{\partial \rr}+T^{-1} \zeta \frac{\pp}{m},\nonumber\\
   -\dot{S} &= T^{-2}\zeta \left(\frac{\pp}{m}\right)^2.
\end{align*}
\end{subequations}
The first equation is clearly unaffected by TRT, it is fully reversible.
The second equation contains a reversible contribution
(the first term of the r.h.s., which transforms as the l.h.s.)
and an irreversible contribution (the second term, sign of which is flipped).
The third equation is fully irreversible (sign flipped).
Hence, TRT is a~mean how to distinguish between reversible and irreversible evolution;
see \cite{PRE15} for more details and for the geometric definition
of the transformation
.

\subsection{Dual Dissipation Potentials}
One can perform
 the Legendre transformation of the dissipation potential $\Xi$
\begin{equation*}
	\Xi^*(\qq,\qq^\circ)
	=
	-\Xi(\qq,\qq^*(\qq^\circ)) + \qq^\circ\cdot\qq^*,
\end{equation*}
The gradient dynamics \eqref{eq.GD} can then be rewritten in terms of $\Xi^*$
\begin{equation}\label{eq.GENERIC.d}
	\frac{\partial \Xi^*}{\partial \qq^\circ}\Big|_{\qq^\circ = \dot{\qq}-\LL\cdot\dd E}  = \frac{\partial S}{\partial \qq^*}
\end{equation}
and one has the liberty to decide which dissipation potential and formulation of GENERIC
to work with,
be it either for physical or mathematical reasons;
see e.g. 
\cite{AmbrosioGigliSavare2008,YangStainerOrtiz2006,Mielke2016,JuengelStefanelliTrussardi2020},
or for nonconvex dissipation potentials also \cite{nonconvex,ContiMuellerOrtiz2020}.

\subsection{Isothermal Case} \label{subsec:Isot}
For temperature $\theta$ as a state variable,
i.e. $\qq = (\tilde{\qq},\theta)$,
the generating potentials are the total Helmholtz free energy
$F(\tilde{\qq},\theta)$ and 
(i.e. including also the kinetic contribution, as opposed to internal Helmholtz free energy)
and free entropy
$P\left(\tilde{\qq},\frac{1}{\theta}\right)$
(sometimes also called Massieu potential),
in the energetic and entropic representation respectively.
In the isothermal case, where $T$ denotes the heat bath temperature,
we define resembling functionals (not thermodynamic potentials in~the~strict sense)
\begin{align} \label{eq:Exergy}
	B(\tilde{\qq},\theta) := E(\tilde{\qq},S(\tilde{\qq},\theta)) - T S(\tilde{\qq},\theta),
\end{align}
which is called exergy (available energy), and
\begin{align}\label{eq:Extropy}
	C\left(\tilde{\qq},\frac{1}{\theta}\right)
	:=
	S\left(\tilde{\qq},E(\tilde{\qq},\frac{1}{\theta})\right)
	- \frac{1}{T} E\left(\tilde{\qq},\frac{1}{\theta}\right),
\end{align}
which satisfy the relations
\begin{align} \label{eq:FreeESRelation}
	C\left(\tilde{\qq},\frac{1}{\theta}\right)
	=
	-\frac{1}{T} B(\tilde{\qq},\theta),
\end{align}
and
\begin{align} \label{eq:ExergyEqui}
	\left. \pder{B}{\theta} \right|_{\theta = T} = 0.
\end{align}
For simplicity we also assume that the dissipation potential $\Xi$ is quadratic.
Then we can use the~single functional $\tilde{F}(\tilde{\qq}) := B(\tilde{\qq},T) = F(\tilde{\qq},T)$
to generate the evolution equations for $\tilde{\qq}$
around the equilibrium temperature $T$.
Indeed, expressing $E$ and $S$ from the relations
\eqref{eq:Exergy} and \eqref{eq:Extropy} respectively,
and plugging them into the~evolution equation \eqref{eq:GENERIC-mat} yields%
\footnote
{
	Here we exploit the simple relation between the~differentials
	of $E$, $S$, $B$, and $C$,
	which is valid since $T$ is a~constant and not~a~spatial dependent field.
	It has to be stressed that for general free energy or entropy
	(depending e.g. on gradients of the fields) in non-isothermal case
	no such naive relation holds;
	see \cite{Mielke2016}.
}%
\begin{align*}
	\dot{q}^i
	&=
	L^{ij} \frac{\partial E}{\partial q^j} + M^{ij}\frac{\partial S}{\partial q^j}\nonumber\\
    &=
	L^{ij} \frac{\partial B}{\partial q^j}
	+
	T \underbrace{L^{ij} \frac{\partial S}{\partial q^j}}_{=0}
	+ M^{ij} \frac{\partial C}{\partial q^j} 
	+ \frac{1}{T} \underbrace{M^{ij} \frac{\partial E}{\partial q^j}}_{=0}
	=
	L^{ij} \frac{\partial B}{\partial q^j}
	- \frac{1}{T} M^{ij} \frac{\partial B}{\partial q^j}
	=
	\left(
		L^{ij}
		- \frac{1}{T} M^{ij}
	\right)
	\frac{\partial B}{\partial q^j}.
\end{align*}
where the degeneracies \eqref{eq:deg-S} and \eqref{eq:deg-E} were employed.
Using the~equilibrium property \eqref{eq:ExergyEqui}
we can neglect for $\theta \approx T$ the last column of $\LL$ and $\MM$,
while the last equation for the temperature $\theta$ is satisfied approximately,
provided that the latent and dissipative heat production are relatively small
compared to the~transfer with the heat bath;
see e.g. \cite[sec. 2.6]{Mielke2011} for more details.
The evolution equations for the reduced state variables $\tilde{\qq}$ hence become
\begin{equation}\label{eq.gen.iso}
	\dot{\tilde{q}}^i
	=
	\tilde{L}^{ij} \frac{\partial F}{\partial \tilde{q}^j}
	- \frac{1}{T} \tilde{M}^{ij} \frac{\partial F}{\partial \tilde{q}^j}
	=
	\tilde{L}^{ij} \frac{\partial \tilde{F}}{\partial \tilde{q}^j}
	- \frac{1}{T} \tilde{M}^{ij} \frac{\partial \tilde{F}}{\partial \tilde{q}^j},
\end{equation}
where $\tilde{L}^{ij}$ and $\tilde{M}^{ij}$ are obtained respectively from $L^{ij}$ and $M^{ij}$
by dropping the last row and column.
In~a~coordinateless form we have
\begin{align}
	\label{eq.GENERIC.isot}
    \dot{D} = \{D,\tilde{F}\}^{\text{iso}} - \frac{1}{T} \d D \cdot \mathbf{\tilde{M}} \cdot \d \tilde{F},
\end{align}
for every functional $D=D(\tilde{\qq})$.

In the isothermal case with a~fixed temperature $T$,
it is hence enough to have the total free energy $\tilde{F}(\tilde{\qq}) = F(\tilde{\qq},T)$,
which then generates both reversible and irreversible evolution.
Plugging it into \eqref{eq.GENERIC.isot} yields the dissipation rate
\begin{align} \label{eq:DissRate}
	-\dot{\tilde{F}}
	=
	- \{\tilde{F},\tilde{F}\}^{\text{iso}} + \frac{1}{T} \d \tilde{F} \cdot \mathbf{\tilde{M}} \cdot \d \tilde{F}
	=
	\frac{2}{T} \Xi
	\geq 0,
\end{align}
where the antisymmetry of the Poisson bracket
and the positive $2$-homogeneity of
the quadratic dissipation potential \eqref{eq:DissPotQuad} were used.
Free energy is thus reduced.
By~integrating
\eqref{eq:DissRate}
from time $t_1$ to $t_2$
one arrives at energy equality
\begin{align*}
	\tilde{F}(t_2) +\frac{1}{T} \int_{t_1}^{t_2} 2 \Xi \, \d t = \tilde{F}(t_1).
\end{align*}
In our toy model, the isothermal case is driven by total free energy
\begin{equation*}
	\tilde{F}(\mathbf{\tilde{q}}) = F(\rr,\pp,T) = \frac{|\pp|^2}{2m} + V(\rr) + \epsilon(S(\rr,\pp,T)) - T S(\rr,\pp,T)
\end{equation*}
and by the same equations as before.
The difference between the isothermal and non-isothermal regime becomes apparent
in the continuum case where also gradients of temperature play a role, see e.g. \cite{dynmaxent}.

\subsection{Summary}
Let us now summarize the construction of GENERIC.
One first needs the set of state variables $\qq$.
Once they are chosen, the Poisson bracket expressing their kinematics is usually known
(typically by a geometric argument).
This is one of the key elements of GENERIC,
to focus on geometric mechanics instead of, for instance, on conservation laws,
which are then rather a~consequence of symmetries in the thermodynamic system.
For a specific energy, one can write the reversible equations in a closed form.
Irreversible evolution is generated by (generalized) gradient dynamics.
The dissipation potential is typically convex in the conjugate variables
and has a minimum at zero.
After conjugate variables are identified with derivatives of the~prescribed entropy,
also, the irreversible evolution gets a closed form.
Complete evolution equations of the state variables $\qq$ are the sum
of the reversible and irreversible contributions.
The evolution of any functional $D(\qq)$ is given by
the General Equation for Non-Equilibrium Reversible and Irreversible Coupling
\begin{equation*} 
    \dot{D}
    =
    \{D,E\}
    +
    \left\langle
      \frac{\delta D}{\delta \qq},
      \frac{\delta \Xi}{\delta \qq^*}\Big|_{\qq^*=\frac{\delta S}{\delta \qq}}
    \right\rangle,
\end{equation*}
in~a~coordinateless form, or by the evolution equations of the state variables \eqref{eq.GENERIC}.
In~the~isothermal case one uses the analogues \eqref{eq.GENERIC.isot} and \eqref{eq.gen.iso}.

\section{Isotropic Model} \label{sec:iso}
Having introduced the fundamental concepts of GENERIC,
we move now
to explaining the basics of NCF on~a~continuum level
and then to
illustrating both frameworks on a simple continuum model for non-Newtonian fluids
-- the isotropic Maxwell, Oldroyd-B, and Giesekus models.
For simplicity, we shall be constrained to isothermal processes.

\subsection{Framework of Natural Configurations}
This framework is suitable for
phenomenological
non-equilibrium continuum thermodynamics;
originally developed in \cite{RajagopalSrinivasa2000},
later refined in \cite{MaRaTu2015,MaRaTu2018} and summarised in \cite{malek.j.prusa.v:derivation}.
Within this framework one introduces a so called \emph{natural configuration}
and assumes that evolution between a reference configuration and the natural configuration is irreversible (or dissipative)
while evolution between the natural configuration and a current configuration is reversible (i.e. purely elastic).
In particular, one can diminish the role of the reference configuration,
which gradually loses its physical importance, for instance due to plastic deformations; \cite{tamas-kinematics}.
This splitting resembles the GENERIC splitting of the evolution equations into mechanics and thermodynamics.
Let us describe the NCF in more detail.

\subsubsection{Balance Laws} \label{subsubsec:BalanceLaws}
The framework builds on the balance equations in continuum mechanics.
We list them in the simplest possible form including only the terms important for our further derivations;
c.f. \cite{GurtinAnand2010}.
We start with the balance of mass that we will use in the standard form
\begin{equation}
	\dot{\rho} = - \rho \dive \vv,
\end{equation}
where $\rho$ is density, $\vv$ fluid velocity and $\dot{\rho} := \frac{\partial \rho}{\partial t} + \vv \cdot \nabla \rho$ denotes hereafter the convective, or material, derivative.
Balance of linear momentum reads
\begin{equation}
	\rho \dot{\vv} = \dive \TTT,
\end{equation}
where $\TTT$ denotes the Cauchy stress tensor; note that
we suppose no external body force.
We will further consider continuum with no internal moment of inertia,
where the only moment interaction is due to the moment of~the~surface forces.
This gives us the balance of angular momentum in the simple, algebraic form
\begin{equation}
	\TTT = \TTT^\top.
\end{equation}
For the internal energy balance, we will not consider any external heat source in the body of our fluid.
Therefore
\begin{equation}\label{bal_energy}
	\rho \dot{e} = \TTT : \L - \dive \mathbf{q},
\end{equation}
where $e$ is the internal energy, and $\mathbf{q}$ is the heat flux and $\L$ denotes the velocity gradient, i.e. $\L:=\nabla\vv$.

\subsubsection{Thermodynamics} \label{subsubsec:NCFThermodynamics}
To~formulate the thermodynamics we rewrite
the~balance of internal
energy in~terms of~entropy $\eta$, which then takes
the~form
\begin{equation}\label{bal_entropy}
	\rho \dot{\eta} = - \dive\mathbf{q}_{\eta} + \zeta,
\end{equation}
where we introduced the entropy flux $\mathbf{q}_{\eta}$ and the entropy production $\zeta$.
The second law of thermodynamics states that $\zeta \geq 0$. 

If we assume that the thermodynamic temperature is constant\footnote{
	In fact the temperature can not be constant because the energy dissipates in the body.
	However, we can assume that either the heat capacity is so large that the temperature changes only negligibly,
	or the body is placed in the big thermal reservoir and the body conducts the heat so fast that the energy is transfered away almost immediately.
},
i.e. $T=const.$,
and moreover that the entropy flux is strictly related to the heat flux
by $\mathbf{q}_{\eta}=\mathbf{q}_{e}/T$,
we can arrive by subtracting $\eqref{bal_energy}-T\eqref{bal_entropy}$
at the reduced thermodynamic identity
\begin{equation}\label{reduced_thermodynamic_identity}
	0\leq\xi:=T \zeta=\TTT : \L - \rho\dot{\psi},
\end{equation}
where $\xi$ is the dissipation rate per volume
and $\psi := e-T\eta$ is the internal Helmholtz free energy density.
The constitutive relation for $\TTT$ and the evolution equations of~the~remaining quantities (i.e. except velocity)
has to~be~such that the~inequality \eqref{reduced_thermodynamic_identity} is satisfied.

\subsubsection{Kinematics of the Natural Configuration} \label{subsubsection:NCFKinematics}
\begin{figure}[h]
\begin{center}
\includegraphics[width=8cm]{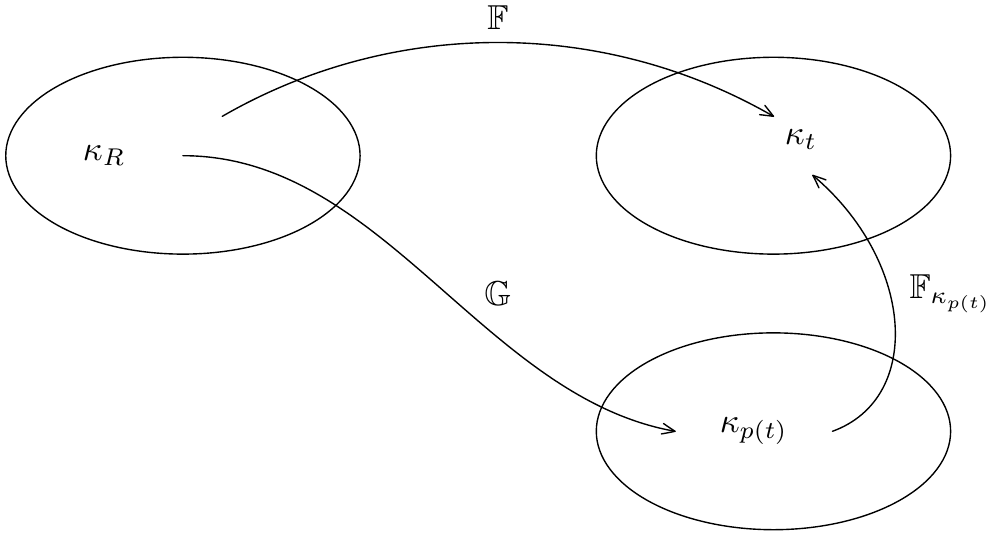}
\end{center}
\caption{Natural configuration.}\label{fig_natural_conf}
\end{figure}

Besides standard reference configuration $\kappa_R$ and current configuration $\kappa_t$, we define a 
natural configuration $\kappa_{p(t)}$, see Figure~\ref{fig_natural_conf}.
It is a configuration of the body associated with the current
configuration $\kappa_t$ at time $t$ that would be obtained if the external stimuli are suddenly removed.
In general, $\kappa_{p(t)}$ does not exist globally.
For more details on this notion see \cite{RajagopalSrinivasa2000}.

Because during the sudden relaxation of the current configuration only the elastic (reversible) part of the 
deformation occurs, we can suppose that the internal free energy of the system is hidden only in the deformation of the natural configuration.
Hence we
split the total deformation gradient $\F$ into the purely elastic reversible
deformation described by $\F_{\kappa_{p(t)}}$
(hereafter just $\F\kpt$ for the sake of brevity)
and irreversible dissipative deformation described by $\mathbb{G}$, such that
\begin{equation} \label{eq:NatConfSplit}
	\F=\Fp\G.
\end{equation}
For description of the kinematics of the natural configuration,
that is defined locally by \eqref{eq:NatConfSplit},
we chose quantities analogical to those in the standard kinematics.
There
the velocity gradient and its symmetric part are related to the deformation gradient $\F$ through
\begin{equation}
\L=\dot{\F}\F^{-1},\quad \D=\frac12(\L+\L^\top).
\end{equation}
Hence we \emph{postulate} the velocity gradient in~the~natural configuration (and its symmetric part) to be
\begin{equation}
\Lp:=\dot{\G}\G^{-1},\quad \Dp:=\frac12(\Lp+\Lp^\top).
\end{equation}
Material time derivative of $\Fp$ and $\Bp$ then reads
\begin{equation}
\dot{\F}\kpt=\L\Fp-\Fp\Lp,\quad \dot{\B}\kpt=\L\Bp+\Bp\L^\top-2\Fp\Dp\Fp^\top.
\end{equation}

\subsubsection{Derivation of Models}
The derivation is based on prescribing two scalar functions.
The first one -- the internal Helmholtz free energy $\psi$ -- is responsible for the elastic part of the response,
the other -- rate of dissipation $\xi$ -- describes how the energy in the body dissipates.
For~the~family of~viscoelastic models, we assume the internal Helmholtz free energy has the form
\begin{equation}
    \psi=\psi(T,\rho,\FFF\kpt)
    =
    \rho \psi_0(T, \rho)
    +
    \rho \psi_\el(T, \FFF\kpt),
\end{equation}
where $T$ is the constant temperature.
Using the reduced thermodynamic identity and the balance of mass we arrive at
\begin{align}
    0 & \leq \xi
    =
    \TTT:\LLL - \rho\dot{\psi}
    =
    \TTT:\LLL - \rho\pder{\psi_0}{\rho}\dot{\rho} - \rho\pder{\psi_\el}{\FFF\kpt}:\dot{\FFF}\kpt \nonumber \\
    &=
    \Big(\TTT + \rho^2\pder{\psi_0}{\rho}\III - \rho\pder{\psi_\el}{\FFF\kpt}\FFF\kpt^\top\Big):\DDD + \rho\pder{\psi_\el}{\FFF\kpt}:\Big(\L\FFF\kpt - \dot{\FFF}\kpt\Big), \label{xi0} \\
    &=
    \Big(\TTT + \rho^2\pder{\psi_0}{\rho}\III - \rho\pder{\psi_\el}{\FFF\kpt}\FFF\kpt^\top\Big):\DDD + \rho\pder{\psi_\el}{\FFF\kpt}:\Fp\Lp,\nonumber
\end{align}
where we used the symmetry of $\TTT + \rho^2\pder{\psi}{\rho}\III - \rho\pder{\psi}{\FFF\kpt}\FFF\kpt^\top$.
The~equations are closed by prescribing a~relation between
$\left\{ \TTT + \rho^2\pder{\psi}{\rho}\III - \rho\pder{\psi}{\FFF\kpt}\FFF\kpt^\top, \LLL\FFF\kpt-\dot{\FFF}\kpt \right\}$
and $\left\{ \DDD, \rho\frac{\partial \psi}{\partial \FFF\kpt} \right\}$,
eventually involving $\rho$ and $\FFF\kpt$,
in such a~way the prescribed rate of dissipation
$\xi$ is met.
Note that this closure is unique when the entropy production is quadratic, but might become more subtle in non-quadratic cases, see \cite{entprodmax}.

\subsubsection{Specific Example}
We repeat the derivation of variant of an isotropic viscoelastic Giesekus rate-type fluid model,
following \cite{MaRaTu2015,MaRaTu2018}.
The full Giesekus model \cite{Giesekus1982} has been derived within the framework of natural configuration in \cite{Dostalik-diffusive_viscoelastic}
and it reads
\begin{align}
\dive\vv&=0,\\
\rho\dot{\vv}&=\dive\TTT,\quad\TTT=-p\I+2\mu\D+G\B,\\
\dot{\B}-\L\B-\B\L^T&=-\frac{1}{\tau}(\alpha\B^2+(1-2\alpha)\B-(1-\alpha)\I),
\end{align}
where $\tau$ is the relaxation time and $\alpha\in(0,1]$. In this paper we present the derivation for $\alpha=1$.
The model uses the dissipation rate function
\begin{equation}
    \xi = 2\mu|\D|^2+\Lambda\Big|\rho\pder{\psi_\el}{\FFF\kpt}\FFF\kpt^\top\Big|^2,
\end{equation}
where $\mu$ and $\Lambda$ are material constants
and $|\mathbb{A}|=\sqrt{A_{ij}A_{ij}}$ is the Frobenius norm.
The first term corresponds to Newtonian viscous dissipation.
The second depends on the elastic Cauchy stress resulting from a~deformation between the~current and the natural configuration,
which is described by $\Fp$;
this term hence gives
the~dissipation caused by~the~evolution of~the~natural configuration.
By closing \eqref{xi0} one obtains
\begin{align}
    \TTT
    &=
    -p_{th}\III + 2\mu\D + \Tel, \nonumber \\
    \dot{\FFF}\kpt
    &=
    \L\FFF\kpt - \Lambda \Tel \FFF\kpt
    \label{giesF1} \\
    \label{gies1}
    \Tel
    &=
    \rho \pder{\psi_\el}{\FFF\kpt}\FFF\kpt^\top,
\end{align}
with the usual formula for pressure
\begin{equation}\label{eq.p.psi}
	p_{th} := \rho^2 \frac{\partial \psi_0}{\partial \rho}.
\end{equation}

To close
the equation \eqref{gies1} we prescribe
\begin{align} \label{neo-Hook}
	\psi_\el(T, \FFF\kpt)
    =
    \frac{G(T)}{2} (|\FFF\kpt|^2 - 3)
    -
    k_B T \ln \det (\FFF\kpt \FFF\kpt^\top),
\end{align}
which is the internal free energy of compressible neo-Hookean solid,
$k_B$ being the Boltzmann constant and $G(T)$ is the elastic modulus;
see for example \cite{Ogden:1997,PavelkaKlikaGrmela2018}.
The important fact here is that the logarithm appearing in the free energy comes from entropy and not from the internal energy;
for derivation of the~entropy~of~elastic dumbbells see Appendix \ref{sec.entropy}.
Differentiating \eqref{neo-Hook} gives
\begin{equation}
    \frac{\partial\psi_\el}{\partial \FFF\kpt}
    =
    G(T)\left( \FFF\kpt - \frac{2k_B T}{G(T)}\FFF\kpt^{-\top} \right),
    \label{derpsi1}
\end{equation}
and then inserting it into \eqref{gies1} yields
\begin{align}
    \Tel
    &=
    \rho G(T)\left( \FFF\kpt\FFF^\top\kpt - \frac{2k_B T}{G(T)}\III \right),
\end{align}
Finally, to obtain the more standard evolution equation for $\BBB\kpt := \FFF\kpt \FFF^\top\kpt$,
we sum $\eqref{giesF1}\cdot \FFF\kpt^\top+\FFF\kpt\cdot \eqref{giesF1}^\top$ and arrive at the Giesekus model
\begin{align*}
    \dot{\rho} &= -\rho \dive \vv, \\
    \rho \dot{\vv}
    &=
    \dive
    \left(
        - p_{th} \III
        + 2 \mu \DDD
        + \Tel
    \right), \\
    \dot{\B}\kpt
    &=
    \LLL \BBB_\kappa
    +
    \BBB_\kappa \LLL^\top
    - 2 \Lambda \Tel \BBB\kpt, \\
    \Tel
    &=
    \rho G(T)\left( \BBB\kpt - \frac{2k_B T}{G(T)}\III \right),
\end{align*}
where we used that both $\Tel$ and $\BBB\kpt$ are symmetric and in our special case commute.

\begin{remark}
	The isotropic Oldroyd-B model is obtained by choosing the rate of dissipation
	\begin{equation}
    	\xi=2\mu|\D|^2+\Lambda\Big|\rho\pder{\psi_\el}{\FFF\kpt}\Big|^2,
	\end{equation}
	depending rather
	on~the~stress tensor between the~natural and the~current configuration.
	Putting the solvent viscosity $\mu=0$ leads to the Maxwell model.
\end{remark}

In summary, once the internal free energy density and entropy production rates are known,
the NCF leads to a system of closed evolution equations for $\rho$, $\vv$, and $\FFF\kpt$,
where the constitutive relation for the Cauchy stress tensor $\TTT$ is known.

\subsection{GENERIC}
Having moved from the~finite dimensional setting to~continuum,
the state variables are no longer elements of $\RRR^n$,
$n \in \NNN$ being the dimension,
but we work with Lagrangian or Eulerian fields instead.
First, we derive the reversible kinematics (Poisson brackets)
in~the Lagrangian and Eulerian frames;
when energy is chosen, the reversible evolution becomes explicit.
Subsequently, we prescribe quadratic dissipation potentials
leading to the well known Maxwell, Oldroyd-B, and Giesekus models.

\subsubsection{Lagrangian Reversible Continuum Mechanics}\label{sec.Lagr}
Let us now generalize the above model of one particle to infinitely many continuum particles, 
where each material point of the continuum has its own label $\XX$ (Lagrangian position).
The actual Eulerian position of the material point in an laboratory frame
is denoted by mapping $\yy(\XX)$.
Compared with the setting of particle mechanics in the Sec. \ref{sec:toy},
this mapping is a continuum analogue of position $\rr$ of the $i$-th particle,
just the index is now continuous.
Therefore, one may anticipate a momentum density field $\MM(\XX)$
being the analogue of $\pp$.
The last state variable would be the entropy field $s(\XX)$,
i.e. its density w.r.t. the volume in the reference configuration;
the total entropy is then given by
\begin{align*}
	S = \int_{\RRR^3} s(\XX) \, \d \XX.
\end{align*}
However, since we want to work in~the~isothermal setting,
we drop it.%
\footnote
{
It is a~matter of~a~straightforward calculation to verify
that such a~shortcut is~compatible with the~definition of~the~isothermal evolution
in~subsection \ref{subsec:Isot},
i.e. the restriction of~the~Poisson bivector $\LL$ to the~variables $(\yy,\MM)$ is not affected
by changing the last state variable from entropy to temperature.
}

\subsubsection{Lagrangian Kinematics}
The analogical Poisson bracket is 
\begin{equation}\label{eq.PB.Lagr}
    \{B, C\}^{\text{Lagrange}}
    =
    \int_{\RRR^3}
        \left(\frac{\delta B}{\delta y^i} \frac{\delta C}{\delta M_i}
        -
        \frac{\delta C}{\delta y^i} \frac{\delta B}{\delta M_i}\right) \,
    \d \XX,
\end{equation}
see e.g. \cite{Goldstein1980,Simo1988,lageul}
for the~definition of~the~bracket and \cite{Morrison1998}
for~an~explanation of~the~functional derivative $\fder{F}{\yy}$
and the~related calculus.
Roughly speaking,
the sum is replaced by an integral (or a suitable duality pairing),
the partial derivatives by functional ones.
This bracket can be alternatively derived from the principle of least action,
having Lagrangian dependent on field $\yy$ and $\dot{\yy}$ in the same fashion as the usual Hamiltonian canonical equations follow from the minimization of action. 

As in the finite dimensional case,
by comparing the both sides of
$\dot{D} = \{D,\tilde{F}\}$
we see that the evolution equations implied by bracket \eqref{eq.PB.Lagr} are
\begin{subequations}
\label{eq.Lagr.evo.gen}
\begin{align}
    \left(\pder{y^i}{t}\right)_{\rev} &= \frac{\delta F}{\delta M_i} \\
    \left(\pder{M_i}{t}\right)_{\rev} &= -\frac{\delta F}{\delta y^i}.
\end{align}
\end{subequations}
This dynamics can be extended to cover also irreversible (plastic) evolution;
see e.g. \cite{Hutter-plastic}.

\subsubsection{Lagrangian Elastic Dynamics}
Typical dependence of energy on fields $\yy$ and $\MM$ is 
\begin{equation*}
    \tilde{F}(\yy,\MM)
    =
    F(\yy,\MM,T)
    =
    \int_{\RRR^3}
        \left(\frac{|\MM|^2}{2\rho_0(\XX)}
        + \rho_0(\XX)
        \psi
        \left(T, \frac{\partial \yy}{\partial \XX} \right)\right) \,
    \d \XX,
\end{equation*}
where $\rho_0(\XX)$ is the field of reference density and $\psi(T,\FFF)$ is the~internal Helmholtz free energy.
The function $W(\FFF) := \rho_0 \psi(T,\FFF)$ is sometimes called the stored elastic energy.
Equations \eqref{eq.Lagr.evo.gen} then become
\begin{subequations}
\label{eq.Lagr.evo}
\begin{align}
    \left(\pder{y^i}{t}\right)_{\rev} &= \frac{M_i}{\rho_0}, \\
    \left(\pder{M_i}{t}\right)_{\rev} &= \frac{\partial}{\partial X^j}\left(\rho_0 \frac{\partial 
        \psi
        }{\partial \frac{\partial y^i}{\partial X^j}}\right)
    = \frac{\partial}{\partial X^j}
    	\left(\frac{\partial W}{\partial \frac{\partial y^i}{\partial X^j}}\right)
    ,
\end{align}
\end{subequations}
where the first equation expresses how the material points move
and the second how momentum of the points is changed.

\subsubsection{Reversible Kinematics of Distortion}
The Lagrangian description, used in Sec. \ref{sec.Lagr}, is very detailed,
we know positions and momenta of every point of the continuum;
however, we aim at Eulerian description of
visco-elastic fluids, where rougher state variables are sufficient.
We may switch from the state variables $(\yy,\MM)$ to $(\rho, \mm, \AAA)$,
the latter being the Eulerian fields of density, momentum, and distortion,
where the last is
defined as the inverse deformation gradient
\begin{align*}
  (\AAA)^I_i = A^I_i := \frac{\partial X^I}{\partial y^i} = \left( \frac{\partial y^i}{\partial X^I} \right)^{-1}.
\end{align*}
The reason for taking the inverse deformation gradient as a new state variable,
and not the deformation gradient itself,
is that it naturally depends on the Eulerian position
in~the~current configuration, which always exists globally;
see e.g. \cite{GodRom2003}.
To draw a~connection to literature we note that
all the models in this paper
involving the distortion can be considered as part of the SHTC framework,
which is compatible with GENERIC as shown in \cite{SHTC-GENERIC}.
Although we consider the distortion as~the more suitable variable,
for~a~better comparison with NCF we then rewrite
the~reversible evolution in~state variables $(\rho,\mm,\Fp)$.
Note that
the~reversible evolution of $\Fp := \AAA^{-1}$ coincides with the~evolution of~the~deformation gradient $\FFF$
as far as
no dissipation is present.
In the final equations, where both the reversible and the irreversible evolution are sum together,
the evolution of $\AAA^{-1}$ coincides with $\Fp$, but differs from the~evolution of~$\FFF$.
Nevertheless, we will still denote the distortion in both cases by $\AAA$.

Let us now move to~the derivation of~the kinematics of~distortion by the~standard method called projection.
The idea is very simple,
just a~mere substitution of functionals dependent on the Lagrangian variables only through the Eulerian variables,
which is a standard way towards Eulerian Poisson brackets.
Abarbanel et al. \cite{abarbanel} used the transformation to derive the bracket of fluid mechanics,
Edwards and Beris \cite{Edwards1991,BE} extended the procedure to viscoelasticity (having the conformation tensor as an Eulerian state variable).
An assumption about curl of the deformation tensor being zero was employed in the latter two works, which we can not afford when working with distortion,
since the distortion can generally have non-zero curl (e.g. in the case of dislocations \cite{landau7}).
Details of the Lagrange $\rightarrow$ Euler transformation without that assumption can be found in \cite{lageul}.
The resulting Poisson bracket bracket for the Eulerian state variables, $(\rho, \mm, \AAA)$,
consists of a `fluid mechanics' part, i.e. the bracket
for density and momentum, and two additional terms for the distortion
\begin{align*}
    \{B, C\}^{\text{A}}
    &=
    \{B,C\}^{\text{FM}} \\
    &
    + 
    \int_{\RRR^3}
        A^I_i
        \left(
            \frac{\partial}{\partial x^j}
            \frac{\delta B}{\delta A^I_j} \frac{\delta C}{\delta m_i}
            -
            \frac{\partial}{\partial x^j}
            \frac{\delta C}{\delta A^I_j} \frac{\delta B}{\delta m_i}
        \right) \,
    \d \xx \\
    &
    +
    \int_{\RRR^3}
        \left(
            \frac{\partial A^I_i}{\partial x_j} - \frac{\partial A^I_j}{\partial x_i}
        \right)
        \left(
            \frac{\delta B}{\delta A^I_j} \frac{\delta C}{\delta m_i}
            -
            \frac{\delta C}{\delta A^I_j} \frac{\delta B}{\delta m_i}
        \right) \,
    \d \xx,
\end{align*}
where $\{B,C\}^{\text{FM}}$ is the Poisson bracket expressing kinematics of fluid mechanics,
\begin{align}\label{eq.PB.FM}
 \{B,C\}^{\text{FM}}
 &=
 \nonumber
 \int_{\RRR^3}
        \rho
        \left(
            \frac{\partial}{\partial x^i}
            \frac{\delta B}{\delta \rho} \frac{\delta C}{\delta m_i}
            -
            \frac{\partial}{\partial x^i}
            \frac{\delta C}{\delta \rho} \frac{\delta B}{\delta m_i}
        \right) \,
    \d \xx \\
    \nonumber
    & \quad
    +
    \int_{\RRR^3}
        m_i
        \left(
            \frac{\partial}{\partial x^j}
            \frac{\delta B}{\delta m_i} \frac{\delta C}{\delta m_j}
            -
            \frac{\partial}{\partial x^j}
            \frac{\delta C}{\delta m_i} \frac{\delta B}{\delta m_j}
        \right) \,
    \d \xx.
\end{align}
Note that this Poisson bracket is no longer canonical, which is typical in Eulerian continuum thermodynamics \cite{Beris1991,grcontmath}.

The same localization procedure as for the Lagrange bracket
leads to evolution equations
\label{eq.rev.A}
\begin{align}
    \left( \frac{\partial \rho}{\partial t} \right)_{\text{rev}}
    &=
    - \frac{\partial}{\partial x^i} \left( \rho \frac{\delta F}{\delta m_i} \right), \\
    \left( \frac{\partial m_i}{\partial t} \right)_{\text{rev}}
    &=
    - \frac{\partial}{\partial x^j} \left( m_i \frac{\delta F}{\delta m_j} \right)
    - \rho \frac{\partial}{\partial x^i} \frac{\delta F}{\delta \rho}
    - m_j \frac{\partial}{\partial x^i} \frac{\delta F}{\delta m_j}
    - A^J_j \frac{\partial}{\partial x^i} \frac{\delta F}{\delta A^J_j} \\
    & \quad
    + \frac{\partial}{\partial x^i} \left( A^J_j \frac{\delta F}{\delta A^J_j} \right)
    + \frac{\partial}{\partial x^j} \left(- A^J_i \frac{\delta F}{\delta A^J_j} \right), \\
    \left( \frac{\partial A^I_i}{\partial t} \right)_{\text{rev}}
    &=
    - \frac{\partial}{\partial x^i} \left( A^I_j \frac{\delta F}{\delta m_j} \right)
    +
    \left( \frac{\partial A^I_j}{\partial x^i}
             - \frac{\partial A^I_i}{\partial x^j}
    \right)
    \frac{\delta F}{\delta m_j}
    =
    -\pder{A^I_i}{x^j} \fder{F}{m_j} - A^I_j \pder{}{x^i} \fder{F}{m_j}.
\end{align}
Note that $v^i = \fder{E}{m_i}$ is the~velocity
and hence fluid mechanics (Euler compressible equations) are obtained when the energy is independent of $\AAA$.
At~the~same time, the~equation for~$\AAA = \FFF^{-1}$ is compatible with the~standard kinematics of~the~deformation gradient
$\dot{\FFF} = \LLL \FFF$.
Last but not least, these equations are valid for any energy functional,
e.g. not only for~the standard energy of~simple fluids,
but also for Korteweg fluids and other;
for more details about this derivation see e.g. \cite{PavelkaKlikaGrmela2018}.
Concrete example of the~free~energy functional and the corresponding explicit evolution equations are given in the next subsection.

\subsubsection{Elastic Dynamics of Distortion}
Assuming the typical form
\begin{align} \label{eq:Helmholtz}
    \tilde{F}(\rho, \mm, \AAA) &= F(\rho, \mm, T, \AAA)
    =
    \int_{\RRR^3}
        f(\rho, \mm, T, \AAA) \,
    \d \xx \\
    &=
    \int_{\RRR^3}
        \left(\frac{|\mm|^2}{2 \rho}
        +
        \rho \psi(T, \rho, \AAA)\right) \,
    \d \xx
    =
    \int_{\RRR^3}
        \left(\frac{|\mm|^2}{2 \rho}
        +
        \rho\psi_0(T, \rho)
        +
        \rho \psi_\el(T, \AAA)\right) \,
    \d \xx,\nonumber
\end{align}
we can now rewrite the evolution equation for momentum density to a more standard form.
Since the density $f$ depends on the fields only in an algebraic manner,
the functional derivatives of $F$ are represented by the partial derivatives of $f$.
A direct computation leads to
\begin{align*}
    \left( \frac{\partial m_i}{\partial t} \right)_{\text{rev}}
    &=
    - \frac{\partial}{\partial x^j} \left( m_i \frac{m_j}{\rho} \right)
    - \rho \frac{\partial}{\partial x^i} \frac{\partial f}{\partial \rho}
    - m_j \frac{\partial}{\partial x^i} \frac{\partial f}{\partial m_j}
	- A^J_j \frac{\partial}{\partial x^i} \frac{\partial f}{\partial A^J_j}
    + \frac{\partial}{\partial x^i} \left( A^J_j \frac{\partial f}{\partial A^J_j} \right)
    - \frac{\partial}{\partial x^j} \left( A^J_i \frac{\partial f}{\partial A^J_j} \right) \\
    & 
    =
    - \frac{\partial}{\partial x^j} \left( m_i \frac{m_j}{\rho} \right)
    - \frac{\partial}{\partial x^i}
    \left(
        - f
        + \rho \frac{\partial f}{\partial \rho}
        + m_j \frac{\partial f}{\partial m_j}
    \right)
    \frac{\partial}{\partial x^j} \left( -A^J_i \frac{\partial f}{\partial A^J_j} \right) \\
    & 
    =
    - \frac{\partial}{\partial x^j} \left( m_i \frac{m_j}{\rho} \right)
    - \frac{\partial}{\partial x^i}
    \left(
        \rho^2 \frac{\partial \psi_0}{\partial \rho}
    \right)
    + \frac{\partial}{\partial x^j} \left( - \rho A^J_i \frac{\partial \psi}{\partial A^J_j} \right).
\end{align*}
Simplifying the other equations for this particular
total free energy ansatz leads to the final system
\begin{align}
	\nonumber
    \left( \frac{\partial \rho}{\partial t} \right)_{\text{rev}}
    &=
    - \dive \left( \rho \vv \right), \\
	\nonumber
    \left( \frac{\partial \mm}{\partial t} \right)_{\text{rev}}
    &=
    - \dive \left( \mm \otimes \vv \right)
    + \dive(-p_{th}\III + \Tel), \\
	\nonumber
    \left( \frac{\partial \AAA}{\partial t} \right)_{\text{rev}}
    &=
    - (\nabla \AAA) \vv
    - \AAA \LLL, \\
    \label{eq:GENERICTelAni}
    \Tel &= -\rho \AAA^\top \frac{\partial \psi_\el}{\partial \AAA}.
\end{align}
where we defined
\begin{align*}
    \vv := \frac{\delta F}{\delta \mm}
    = \frac{\partial f}{\partial \mm} = \frac{\mm}{\rho},
    \quad
    p_{th}
    := \rho^2 \frac{\partial \psi}{\partial \rho}
    = \rho^2 \frac{\partial \psi_0}{\partial \rho}.
\end{align*}
Rewriting these equations in~terms of~$(\rho,\vv,\Fp)$ and the using the~convective time derivative gives
\begin{subequations}
\begin{align*}
    \dot{\rho}
    &=
    - \rho \dive \vv, \\
    \rho \dot{\vv}
    &=
    \dive (-p_{th}\III + \Tel), \\
    \dot{\FFF}\kpt
    &=
    \LLL \Fp, \\
    \Tel &= \rho \frac{\partial \psi_\el}{\partial \Fp} \Fp^\top.
\end{align*}
\end{subequations}

For closing the~part of~momentum equation denoted by $\Tel$
we have to specify the energy $\psi_\el$.
A~suitable choice for suspensions of polymeric dumbbells is 
\begin{align} \label{eq:HelmholtzA}
    \psi_\el(T,\AAA)
    =
    \frac{G(T)}{2} (|\AAA^{-1}|^2 - 3)
    -
    k_B T \ln \det (\AAA^{-1} \AAA^{-\top}),
\end{align}
or equivalently
\begin{align} \label{eq:HelmholtzF}
    \psi_\el(T,\Fp)
    =
    \frac{G(T)}{2} (|\Fp|^2 - 3)
    -
    k_B T \ln \det (\Fp \Fp^{\top}),
\end{align}
as it can be derived by methods of statistical physics, e.g. \cite{Carreau-Grmela,PavelkaKlikaGrmela2018} or Appendix \ref{sec.entropy}.
This energy leads to the elastic Cauchy stress tensor \eqref{eq:GENERICTelAni}
\begin{align}
	\label{eq:ElCauchyIso}
    \Tel
    &=
    \rho G(T) \left( \AAA^{-1}\AAA^{-\top} - \frac{2 k_B T}{G(T)} \III \right)
    =
    \rho G(T) \left( \Fp\Fp^{\top} - \frac{2 k_B T}{G(T)} \III \right),
\end{align}
which is nothing but the standard compressible neo-Hookean model as $\BBB\kpt = \AAA^{-1}\AAA^{-\top} = \Fp \Fp^\top$.

\begin{remark}
    Note that there is no material or phenomenological constant in front of the logarithm,
    only the universal Boltzmann constant $k_B$.
    It is therefore straightforward to see
    how the internal free energy should look for the non-isothermal setting.
\end{remark}

\subsubsection{Irreversible Evolution of Distortion}
Since we work in~the~isothermal setting,
the~localized irreversible evolution in~coordinates is given by~the~corresponding part in \eqref{eq.gen.iso}.
For~$\tilde{\qq} = (\rho,\mm,\AAA)$ and quadratic dissipation potential \eqref{eq:DissPotQuad} this becomes
\begin{equation}
	\label{eq:EulerIrr}
	\begin{split}
		\left( \frac{\partial \rho}{\partial t} \right)_{\text{irr}}
		&=
		-\frac{1}{T}
		\left.
		    \frac{\delta \Xi}{\delta \rho^*}
		\right|_{\rho^* = \frac{\delta F}{\delta \rho}}, \\
		\left( \frac{\partial m_i}{\partial t} \right)_{\text{irr}}
		&=
		-\frac{1}{T}
		\left.
		    \frac{\delta \Xi}{\delta (m^*)^i}
		\right|_{(m^*)^i = \frac{\delta F}{\delta m_i}} , \\
		\left( \frac{\partial A^I_i}{\partial t} \right)_{\text{irr}}
		&=
		-\frac{1}{T}
		\left.
		    \frac{\delta \Xi}{\delta (A^*)^i_I}
		\right|_{(A^*)^i_I = \frac{\delta F}{\delta A^I_i}}.
	\end{split}
\end{equation}
For~the~equivalent choice $\tilde{\qq} = (\rho,\mm,\Fp)$ the~last equation is replaced by
\begin{equation}
	\label{eq:Firr}
	\left( \frac{\partial \Fp}{\partial t} \right)_{\text{irr}}
	=
	-\frac{1}{T}
	\left.
	    \frac{\delta \Xi}{\delta \Fp^*}
	\right|_{\Fp^* = \frac{\delta F}{\delta \Fp}}.
\end{equation}
Note that gradient dynamics is invariant with respect to transformations of state variables, see e.g. \cite{JSP2020}.
Apart from the standard properties of $\Xi$ summarised in the subsection \ref{subsec:toy-irr},
the dissipation potential has to be such that
the total mass $M := \int_{\RRR^3} \rho \, \d \xx$ is conserved.
This condition is for example satisfied when $\Xi$ depends merely on gradients of $\rho^*$.
Here we suppose $\Xi$ does not depend on~$\rho^*$ at~all.
Momentum conservation can be ensured by the analogical condition,
but in the non-isothermal setting one gets a slightly more complex structure due to the coupling with energy dissipation, see \cite{PavelkaKlikaGrmela2018}.

\subsubsection{Dissipative Dynamics of Distortion}
\label{subsec:iso-diss}
Let us now show how one can derive a variant
of an isothermal Giesekus model of viscoelastic fluids.
As in the elastic case, we will specify the total Helmholtz free energy,
which is more suitable for isothermal processes.
The conjugate variables,
either $(\mm^*,\AAA^*)$ or $(\mm^*,\Fp^*)$,
are then replaced by the corresponding derivatives of total free energy.
Using \eqref{eq:Helmholtz},
we make the substitution 
\begin{align*}
    \mm^* &\rightarrow
    \frac{\partial f}{\partial \mm} = \frac{\mm}{\rho}, \\
    \AAA^* &\rightarrow
    \frac{\partial f}{\partial \AAA} = \rho \frac{\partial \psi_\el}{\partial \AAA}, \\
    \Fp^* &\rightarrow
    \frac{\partial f}{\partial \Fp} = \rho \frac{\partial \psi_\el}{\partial \Fp}.
\end{align*}
In the isothermal setting the choice of dissipation potential is simply
\begin{align}
    \Xi(\AAA,\mm^*,\AAA^*)
    &=
    T
    \int_{\RRR^3}
        \mu \left|(\nabla \mm^*)_{sym} \right|^2
        +
        \frac{\Lambda}{2} \left| \AAA^\top \AAA^* \right|^2 \,
    \d \xx
    =
    T
    \int_{\RRR^3}
        \mu \left|\DDD \right|^2
        +
        \frac{\Lambda}{2} \left| \rho \AAA^\top \frac{\partial \psi_\el}{\partial \AAA} \right|^2 \,
    \d \xx,
	\label{eq:DissPotA}
\end{align}
or equivalently
\begin{align}
    \Xi(\Fp,\mm^*,\Fp^*)
    &=
    T
    \int_{\RRR^3}
        \mu \left|(\nabla \mm^*)_{sym} \right|^2
        +
        \frac{\Lambda}{2} \left| \Fp^* \Fp^\top \right|^2 \,
    \d \xx
    =
    T
    \int_{\RRR^3}
        \mu \left|\DDD \right|^2
        +
        \frac{\Lambda}{2} \left| \rho \frac{\partial \psi_\el}{\partial \Fp} \Fp^\top \right|^2 \,
    \d \xx,
	\label{eq:DissPotF}
\end{align}
where $\DDD$ stands for the symmetric velocity gradient\footnote{
For a non-isothermal variant see \cite{PavelkaKlikaGrmela2018}.}.
The first term yields the standard viscous dissipation,
the second is chosen as being proportional to the elastic Cauchy stress.
Note also that since the dissipation potential is homogeneous of degree $2$,
it is half the dissipation rate, and prescribing the dissipation potential is equivalent to prescribing the dissipation rate.
It should be also noted that the dissipation drives the system
towards the stress-free configuration if 
the dissipation potential
depends on some stress tensor.
The derivatives of $\Xi$ are
\begin{align*}
    \frac{\delta \Xi}{\delta \mm^*}
    &=
    -T \dive(2 \mu  (\nabla \mm^*)_{sym})
    = - T \dive(2 \mu \DDD), \\
    \frac{\delta \Xi}{\delta \AAA^*}
    &=
    T \Lambda \AAA \AAA^\top \AAA^*
    =
    -T \Lambda \AAA \left( - \rho \AAA^\top \frac{\partial \psi_\el}{\partial \AAA} \right), \\
    \frac{\delta \Xi}{\delta \Fp^*}
    &=
    T \Lambda \Fp^* \Fp^\top \Fp
    =
    T \Lambda \left( \rho \frac{\partial \psi_\el}{\partial \Fp} \Fp^\top \right) \Fp,
\end{align*}
which yields the irreversible evolution
\begin{align*}
    \left(\frac{\partial \mm}{\partial t}\right)_{\irr}
    &=
    \dive(2 \mu \DDD), \\
    \left(\frac{\partial \AAA}{\partial t}\right)_{\irr}
    &=
    \Lambda \AAA \Tel, \\
    \left(\frac{\partial \Fp}{\partial t}\right)_{\irr}
    &=
    - \Lambda \Tel \Fp, \\
    \Tel
    &=
    -\rho \AAA^\top \frac{\partial \psi_\el}{\partial \AAA}
    =
    \rho \frac{\partial \psi_\el}{\partial \Fp} \Fp^\top.
\end{align*}
In particular, choosing $\psi_\el$ as in \eqref{eq:HelmholtzA} and \eqref{eq:HelmholtzF} yields $\Tel$ from \eqref{eq:ElCauchyIso}.

\subsubsection{Final Equations}
Let the state variables be represented by $(\rho,\mm,\AAA)$.
Their reversible evolution has been specified by the Poisson bracket,
leading to equations \eqref{eq.rev.A}.
Once the total free energy is specified (in the isothermal case), as in Eq. \eqref{eq:Helmholtz},
the reversible evolution can be written down explicitly.
The irreversible evolution is given by dissipation potential,
e.g. the one in Eq. \eqref{eq:DissPotA},
and becomes explicit when the specific formula for
free energy (or entropy in the non-isothermal case) is invoked.
The final evolution equations are then the sum of the reversible and irreversible parts of the evolution:
\begin{subequations}
\label{kinnat}
\begin{align}
    \frac{\partial \rho}{\partial t} + \dive (\rho \vv) &= 0, \\
    \frac{\partial \mm}{\partial t} + \dive(\mm \otimes \vv)
    &=
    \dive
    \left(
        - p_{th} \III + 2 \mu \DDD + \Tel
    \right), \\
    \frac{\partial \AAA}{\partial t} + (\nabla \AAA) \vv
    &=
    - \AAA \LLL + \Lambda \AAA \Tel, \\
    \Tel &= - \rho \AAA^\top \frac{\partial \psi_\el}{\partial \AAA},
\end{align}
which are compatible with literature\cite{Dumbser-unified}.
\end{subequations}
Using the~convective time derivative and $\FFF_\kappa := \AAA^{-1}$ instead of the distortion, they can be rewritten to
\begin{subequations}
\begin{align*}
    \dot{\rho} &= -\rho \dive \vv, \\
    \rho \dot{\vv}
    &=
    \dive
    \left(
        -p_{th} \III + 2 \mu \DDD + \Tel
    \right), \\
    \dot{\FFF}\kpt
    &=
    \LLL \FFF_\kappa - \Lambda \Tel \FFF_\kappa, \\
    \Tel &= \rho \frac{\partial \psi_\el}{\partial \FFF_\kappa}\FFF_\kappa^\top,
\end{align*}
\end{subequations}
and in terms of~the left Cauchy--Green tensor $\BBB := \AAA^{-1} \AAA^{-\top} = \FFF_\kappa \FFF_\kappa^\top$
they become
\begin{subequations}
\begin{align*}
    \dot{\rho} &= -\rho \dive \vv, \\
    \rho \dot{\vv}
    &=
    \dive
    \left(
        - p_{th} \III + 2 \mu \DDD + \Tel
    \right), \\
    \dot{\BBB}\kpt
    &=
    \LLL \BBB_\kappa
    +
    \BBB_\kappa \LLL^\top
    - 2 \Lambda \Tel \BBB_\kappa, \\
    \Tel &= 2 \rho \frac{\partial \psi_\el}{\partial \BBB_\kappa}\BBB_\kappa,
\end{align*}
\end{subequations}
where we can easily recognize the upper-convective derivative in the equation for $\BBB$.
This is compatible with the
standard
results\cite{miroslav-elastic}.

If we use the concrete energy \eqref{eq:HelmholtzA},
now written in terms of $\BBB_\kappa$ as
\begin{align*}
    \psi_\el(T,\BBB_\kappa)
    =
    \frac{G(T)}{2} (\tr \BBB_\kappa - 3)
    -
    k_B T \ln \det (\BBB_\kappa),
\end{align*}
we recover a variant of the standard compressible Giesekus model (\cite{Giesekus1982})
\begin{subequations}
\begin{align*}
    \dot{\rho} &= -\rho \dive \vv, \\
    \rho \dot{\vv}
    &=
    \dive
    \left(
        -p_{th} \III
        + 2 \mu \DDD
        + \Tel
    \right), \\
    \dot{\BBB}\kpt
    &=
    \LLL \BBB_\kappa
    +
    \BBB_\kappa \LLL^\top
    - 2 \Lambda \Tel \BBB_\kappa, \\
    \Tel &= 2 \rho G(T)
            \left(
                \BBB_\kappa - \frac{2 k_B T}{G(T)} \III
            \right).
\end{align*}
\end{subequations}

\begin{remark}
	If, instead of dissipation potential \eqref{eq:DissPotA} and \eqref{eq:DissPotF}, we chose
	\begin{align*}
		\Xi(\AAA,\mm^*,\AAA^*)
		&=
		T \int_{\RRR^3}
		    \mu \left|(\nabla \mm^*)_{sym} \right|^2
		    +
		    \frac{\Lambda}{2} \left| \AAA^\top \AAA^* \AAA^\top \right|^2 \,
		\d \xx
		=
		T \int_{\RRR^3}
		    \mu \left|\DDD \right|^2
		    +
		    \frac{\Lambda}{2} \left| \rho \AAA^\top \frac{\partial \psi_\el}{\partial \AAA} \AAA^\top \right|^2 \,
		\d \xx, \nonumber \\
		\Xi(\mm^*,\Fp^*)
		&=
		T \int_{\RRR^3}
		    \mu \left|(\nabla \mm^*)_{sym} \right|^2
		    +
		    \frac{\Lambda}{2} \left| \Fp^* \right|^2 \,
		\d \xx
		=
		T \int_{\RRR^3}
		    \mu \left|\DDD \right|^2
		    +
		    \frac{\Lambda}{2} \left| \rho \frac{\partial \psi_\el}{\partial \Fp} \right|^2 \,
		\d \xx, \nonumber
	\end{align*}
	i.e. the potentials are quadratic in~a~different stress tensor,
	we would obtain the Oldroyd-B model.
	Dropping the viscous part would yield the Maxwell model.
\end{remark}

Let us now proceed to a simple numerical illustration of the dynamics involving distortion.

\subsubsection{Numerical simulation}

\begin{figure}[h]
\begin{center}
\includegraphics[scale=2]{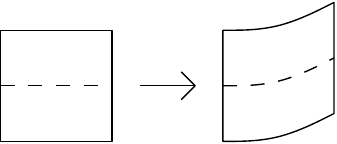}
\caption{Scheme of the simulated vertical simple shear showing the initial configuration (left) and the deformed current configuration (right). Only the deformation of one cut through the material (the dashed line) is examined.}\label{obr0}
\end{center}
\end{figure}

To illustrate equations \eqref{kinnat} in the context of NCF,
we
will show evolution of the natural, the current, the reference configurations.
Let us assume a vertical simple shear between two planes with a distance of one meter.
The reference and current configurations are schematically presented in Figure \ref{obr0}.
We compute the deformation of the dashed line, that is horizontal in the reference configuration with the $x$-coordinates between 0 and 1. 

    Energy and dissipation potential (describing von Mieses plasticity) are taken from \cite{master} and the numerical code is based on the one created and described by \cite{disser}. The initial conditions are set as a linear velocity profile with $v_{x=0}=0 \ \mathrm{m/s}$ and $v_{x=1}=1 \ \mathrm{m/s}$, where $v$ is the Eulerian horizontal velocity. The boundary conditions are set as no-slip to walls at speeds $v_{x=0}=0 \ \mathrm{m/s}$ and $v_{x=1}=1 \ \mathrm{m/s}$.

\begin{figure}
\begin{center}
\includegraphics[width=\columnwidth]{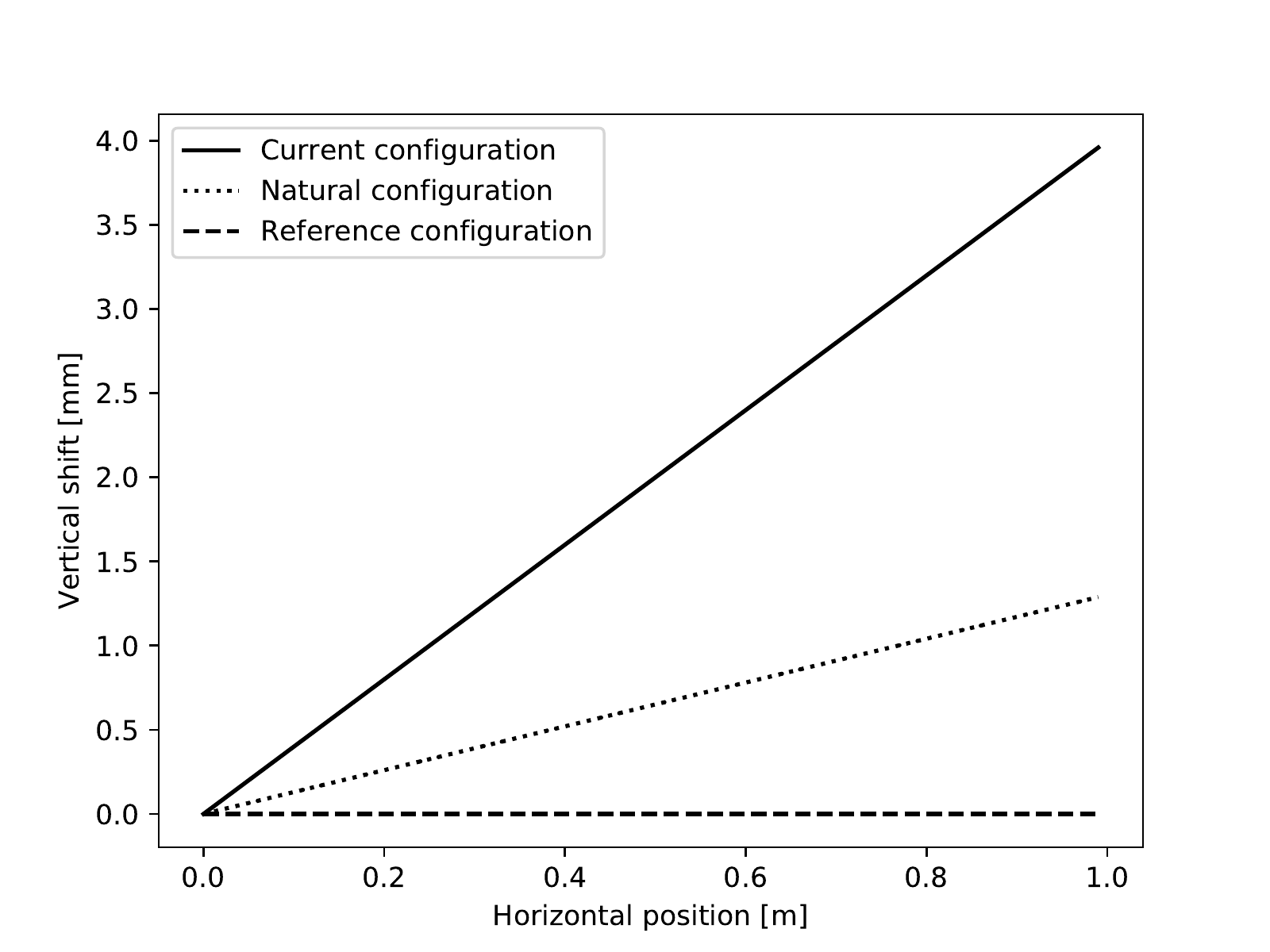}
\caption{Comparison of the reference (dashed), current (solid) and natural (dotted) configurations. Even in such a simple geometry, integration of the distortion matrix $\mathbb{A}$ over the current configuration does not restore the reference configuration.}\label{obr1}
\end{center}
\end{figure}

\begin{figure}
\begin{center}
\includegraphics[width=\columnwidth]{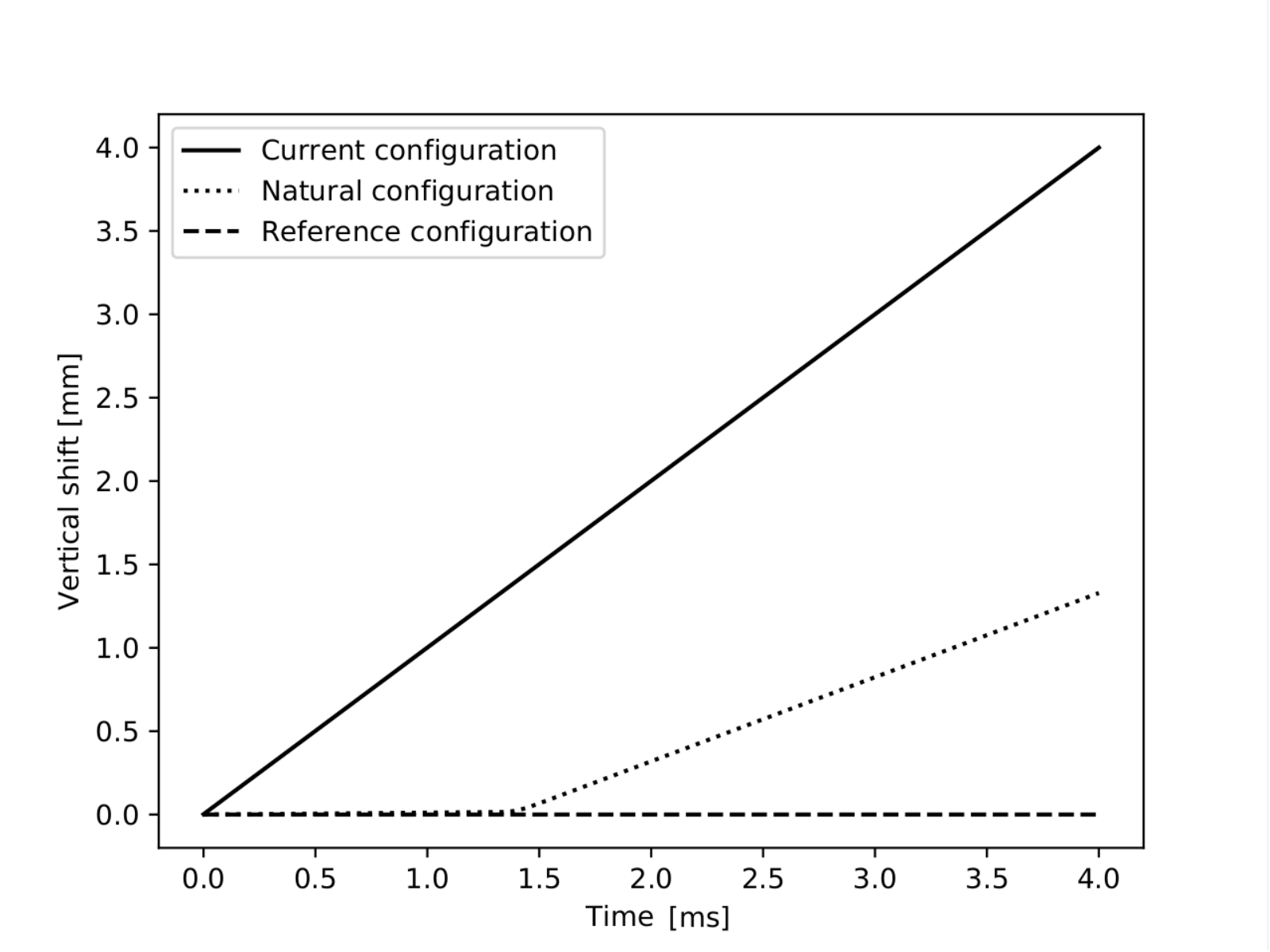}
\caption{Time evolutions of of the shift of the right-hand edge of the body in the reference (dashed), current (solid) and natural (dotted) configurations. For small times, i.e. for small deformations and stresses, the natural and reference configuration coincide. This means that the computed $\mathbb{A}$ is actually the inverse of the deformation gradient. The motion is thus reversible. For larger deformations, the motion becomes irreversible, since the two configurations diverge.}\label{obr2}
\end{center}
\end{figure}

Figure \ref{obr1} shows the vertical shift of the three resulting configurations at time $t=4 \ \mathrm{ms}$.
It can be observed that the three lines are linear,
thus they are fully described by the vertical shift at the right edge of the body, i.e. at $x=1$.
Evolution of the vertical shift in time is plotted in  Figure \ref{obr2}.
One observes that for a short time and small deformations the natural and reference configurations are indistinguishable.
This means the dissipation is negligible and the motion is thus reversible.
After certain threshold is exceeded, the dissipation is switched on,
the motion becomes irreversible and the natural and reference configurations diverge.
This behavior is typical for von Mieses plasticity.
One can thus conclude that the configuration obtained by integrating the distortion matrix $\mathbb{A} = \FFF\kpt^{-1}$
over the current configuration differs from the reference configuration even in a very simple geometry, as long as the dissipation becomes evident.

Therefore, we can start with the current configuration and integrate the distortion to obtain the natural configuration. This is because the deformation between those two configurations is elastic. However, we can not obtain the reference configuration, which has been already lost due to the dissipation. This is the interpretation of NCF within GENERIC and SHTC.

\subsection{Summary}
In NCF one starts with the balance equations for mass, momentum, energy and entropy.
Then three configurations are taken into account:
the reference configuration ($\RRRR$),
the natural configuration ($\NNNN$),
and the current configuration ($\CCCC$),
where the latter coincides with the laboratory frame of reference and Eulerian description.
The dynamics between $\RRRR$ and $\NNNN$ is dissipative,
while from $\NNNN$ to $\CCCC$ purely elastic.
Moreover, the kinematics of the deformation tensor $\FFF$ from $\RRRR$ to $\CCCC$ is the standard, reversible Eulerian kinematics $\dot{\FFF} = \LLL \FFF$.
The irreversible evolution from $\RRRR$ to $\NNNN$ is generated by imposing a dissipation rate
and by choosing constitutive relations leading to~this~rate.
This choice can be~done in a compatible way with GENERIC.
The final evolution observed in $\CCCC$ can be seen as a composition of reversible and irreversible evolution. 

In GENERIC, one starts with the choice of state variables.
The choice can be made equivalent to the natural configuration in NCF, that is $(\rho,\mm,\FFF)$
or a transformation of that, e.g. $(\rho,\mm,\AAA)$.
Then the reversible evolution of the state variables is the usual mechanical evolution,
which is compatible with the evolution from $\NNNN$ to $\CCCC$ in the NCF for $\xi = 0$.
Irreversible evolution is then added both to the equation for momentum density and for the deformation gradient
(or distortion or the left Cauchy--Green tensor).
When the dissipation potential is taken compatible with the dissipation rate in NCF,
the same irreversible evolution is obtained as in NCF.

Let us highlight the similarities in~more detail.
First we explain why the kinematics of~the~distortion
\begin{align*}
	\frac{\partial \AAA}{\partial t}
	+
	(\nabla \AAA)\vv
	=
	- \AAA \LLL - \frac{1}{T} \left. \frac{\delta \Xi}{\delta \AAA^*} \right|_{\AAA^* = \frac{\delta F}{\delta \AAA}},
\end{align*}
is precisely the~kinematics of the natural configuration,
where the choice of $E = E(\AAA)$ and $\Xi = \Xi(\AAA,\AAA^*)$
specifies respectively its elastic and dissipative dynamics.
The reasons are the following:
\begin{itemize}
	\item The reference configuration (which is by definition
	      obtained by tracking the particles' trajectories)
	      can be reconstructed by the integration of a tensor field,
	      which is a solution to the evolution equations
	      \begin{align*}
	          \frac{\partial \AAA}{\partial t}
	            +
	            (\nabla \AAA)\vv
	            =
	            -\AAA \LLL.
	      \end{align*}
	      The complete GENERIC equation for distortion,
	      containing also the irreversible part,
	      yields, in general, a~different solution.
	      This justifies the standard illustrative splitting
	      of the reference, local natural, and current configuration.
	      Moreover, if the tensor field $\AAA$ satisfies
	      the integrability condition, i.e. the~curl of $\AAA$ is zero and the domain is simply connected, the natural configuration
	      exists globally.
	\item Energy depends on the distortion $\AAA$,
	      i.e. the deformation from the natural configuration
	      to the current one is elastic.
	      The Helmholtz free energy typically satisfies
	      \begin{align*}
	          \frac{\partial \psi_{\el}}{\partial \AAA} (T,\III) = 0,
	      \end{align*}
	      i.e. when the material occupies the natural configuration
	      (the natural and current configuration coincides),
	      it is in a~(partial-)stress-free state.
	      In~general, as it is e.g. for $\psi_\el$ from \eqref{eq:HelmholtzA},
	      the~extremum point of~the~free energy is moved to $c \III$ for some $c>0$.
	      The natural configuration, defined by $\AAA$,
	      is then stress-free only up~to~a~homogeneous compression/expansion,
	      leaving non-zero only the~spherical part of~the~stress.
	\item The dissipation potential $\Xi$ is a~function of $\AAA^*$ and we substitute
	      \begin{align*}
	          \AAA^* \to \rho \frac{\partial \psi_\el}{\partial \AAA},
	      \end{align*}
	      i.e. the partial dissipation rate
	      $\frac{1}{T}\AAA^*:\fder{\Xi}{\AAA^*}$,
	      c.f. \eqref{eq:DissRate}, depends
	      on some specific stress measure.
	      This might be seen as an analogue to \cite[Fig. 4]{MalekPrusa2016},
	      where the dissipation is due to the `dashpot' pulled by the `elastic spring'.
	      The dissipation rate
	      can be either expressed in terms of the~`stress in~the elastic spring' in~the~primal formulation using the~dissipation potential $\Xi$,
	      or in~terms of~the~`rate of the dashpot' using the dual dissipation potential $\Xi^*$.
\end{itemize}

The second point we would like to elaborate on is
the~modeling in~both frameworks.
Let us start with plugging the total Helmholtz free energy \eqref{eq:Helmholtz}
into the~equation \eqref{eq:DissRate}, which yields the dissipation rate
\begin{align*}
	-\dot{\tilde{F}}
	&=
	\frac{1}{T}
	\int_{\RRR^3}
		\left(\frac{\delta F}{\delta \mm} : \left. \frac{\delta \Xi}{\delta \mm^*} \right|_{\mm^* = \frac{\delta F}{\delta \mm}}
		+
		\frac{\delta F}{\delta \Fp} : \left. \frac{\delta \Xi}{\delta \Fp^*} \right|_{\Fp^* = \frac{\delta F}{\delta \Fp}}\right) \,
	\d \xx \\
	&=
	\frac{1}{T}
	\int_{\RRR^3}
		\left(\DDD : \left. \frac{\delta \Xi}{\delta \nablasym \mm^*} \right|_{\mm^* = \frac{\delta F}{\delta \mm}} \right. 
		\left.
		+
		\rho \frac{\partial \psi}{\partial \FFF\kpt} : \left. \frac{\delta \Xi}{\delta \Fp^*} \right|_{\Fp^* = \frac{\delta F}{\delta \Fp}}\right) \,
	\d \xx,
\end{align*}
where we supposed $\Xi$ is independent of~$\rho^*$
and depends on~$\mm^*$ merely through its~symmetric gradient.
Defining
\begin{align*}
	\Tirr :&= \frac{1}{T} \frac{\delta \Xi}{\delta \DDD},
\end{align*}
and using the localized evolution for $\Fp$
\begin{align*}
	-\frac{1}{T} \frac{\delta \Xi}{\delta \FFF\kpt^*}
	&=
	\left( \pder{\FFF\kpt}{t} \right)_{\text{irr}}
	=
	\pder{\FFF\kpt}{t} - \left( \pder{\FFF\kpt}{t} \right)_\rev
	=
	\dot{\FFF}\kpt - \LLL \Fp,
\end{align*}
the dissipation rate becomes
\begin{align*}
	-\dot{\tilde{F}}
	&=
	\frac{1}{T}
	\int_{\RRR^3}
		\left(\DDD : \frac{\delta \Xi}{\delta \DDD}
		+
		\rho \frac{\partial \psi_\el}{\partial \FFF\kpt} : \frac{\delta \Xi}{\delta \FFF\kpt^*} \right) \,
	\d \xx
	=
	\int_{\RRR^3}
		\left(\DDD : \TTT_{\text{irr}}
		+
		\rho \frac{\partial \psi_\el}{\partial \FFF\kpt} : (\LLL \Fp - \dot{\FFF}\kpt)\right) \,
	\d \xx.
\end{align*}
On the other hand, the chain rule, the final evolution equations, and integration by parts give
\begin{align*}
	-\dot{\tilde{F}}
	&=
	\left\langle -\frac{\delta F}{\delta \qq}, \partial_t \qq \right\rangle
	=
	\int_{\RRR^3}
		\left(-\vv \cdot \partial_t \mm - (\psi - \frac{1}{2}|\vv|^2) \partial_t \rho - \rho \pder{\psi}{\rho} \partial_t \rho - \rho \pder{\psi}{\Fp} : \partial_t \Fp\right) \,
	\d \xx \\
	&=
	\int_{\RRR^3}
		\left(\vv \cdot \dive(\rho\vv \otimes \vv)
		-\vv \cdot \dive \TTT
		+ (\psi - \frac{1}{2}|\vv|^2) \dive(\rho \vv)
		- \rho \partial_t \psi\right) \,
	\d \xx \\
	&=
	\int_{\RRR^3}
		\left(- \L : (\rho \vv \otimes \vv)
		+ \L : \TTT
		+ \frac{1}{2} \nabla (\vv\cdot\vv) \cdot \rho \vv
		- \rho \dot{\psi}\right) \,
	\d \xx
	=
	\int_{\RRR^3}
		\left(\L : \TTT - \rho \dot{\psi}\right) \,
	\d \xx,
\end{align*}
where again the fact $\left( \pder{\rho}{t} \right)_\irr = 0$ was used.
Hence, by using the positive $2$-homogeneity of~$\Xi$, as~in~\eqref{eq:DissRate},
we obtain by expressing $\dot{\tilde{F}}$ in the two ways above
\begin{align*}
	\frac{2}{T} \Xi = -\dot{F} = \int_{\RRR^3} \xi \, \d \xx,
\end{align*}
which relates the~prescription of~the~dissipation potential $\Xi$ and the~dissipation function $\xi$.
At~the~same time we see that
\begin{align*}
	\int_{\RRR^3} \xi \, \d \xx
	&=
	\int_{\RRR^3}
		\left(\L : \TTT - \rho \dot{\psi}\right) \,
	\d \xx
	=
	\int_{\RRR^3}
		\left(\DDD : \TTT_{\text{irr}}
		-
		\rho \frac{\partial \psi_\el}{\partial \FFF\kpt} : (\LLL \Fp - \dot{\FFF}\kpt)\right) \,
	\d \xx \\
	&=
	\frac{1}{T}
	\int_{\RRR^3}
		\left(\DDD : \frac{\delta \Xi}{\delta \DDD}
		+
		\rho \frac{\partial \psi}{\partial \FFF\kpt} : \frac{\delta \Xi}{\delta \FFF\kpt^*}\right) \,
	\d \xx.
\end{align*}
This equality shows how the~dissipation rate in~NCF has to be rewritten
if one wants to recover the~irreversible evolution in~GENERIC,
and actually explains why the form in~\eqref{xi0} was chosen;
it adopts the~structure given by~the~dissipation potential.
The frameworks are thus compatible with each other,
as was explicitly demonstrated on the isotropic isothermal Giesekus model. 

Finally,
the distortion $\AAA$ can be interpreted as the gradient of the mapping from the current configuration to the natural configuration.
The reference configuration has been ``forgotten'' by the model due to the dissipation,
and can be recovered only by tracking the particles' trajectories.
GENERIC and SHTC thus can provide interpretation of NCF.

\section{Anisotropic Model} \label{sec:aniso}

To further compare the two approaches, we go beyond the isotropic models and include a certain class of anisotropic materials.
By~anisotropy we mean that
material properties are not invariant with respect to rotations.
Materials can exhibit anisotropy with regard to mechanical response in several ways.
For instance, in crystal plasticity, we can have anisotropy with respect to the elastic response,
described by the stored energy function.
We can also have anisotropy associated with the yield surface
and this is captured by the properties of the dissipation potential.
If we modeled liquid crystals, regarded as a rod like suspension in an isotropic fluid,
we would also observe two kinds of anisotropic response, both of a different nature.
Elastic rods lead to an anisotropy in the elastic response,
while the movement of the rods in the viscous fluid, itself isotropic,
causes that the rate of dissipation is different for motions along different directions.
The last example of anisotropy is related to temperature expansion,
which may occur for example in a steel reinforced concrete.
For~anisotropic models for nematic polymers we refer to
\cite{ericksen-anisotropic, Rajabian2005, BE, BarchiesiDeSimone2015, BarchiesiEtAl2017},
for~biological tissues modeling see~\cite{BalzaniNeff2006},
for~modeling of~calendered rubber sheets see~\cite{ItskovAksel2004},
and for~crystal plasticity models we refer to~\cite{Piotr-plasticity}.
We also refer to~\cite{Neff2003,SchroederNeff2003,SchroederNeff2006} for~examples of~anisotropic large-strain energies.

Mathematically speaking,
anisotropy is characterized by a group of transformations with respect to which the material properties are invariant.
In \cite{lokhin-sedov,GurtinAnand2010,ZhangRychlewski1990} it was shown that each of the crystallographic groups can be represented by a set of structural tensors, extra variables.
He we choose perhaps the simplest case when each material point is equipped with a vector $\fibR$ characterizing for instance orientation of a fiber
or a~lattice vector.
In the current configuration (or the laboratory Eulerian frame) the field of the vectors is denoted by $\fibC$.\footnote
{
	Instead of just one vector $\fibC$,
	three vectors seem to be general for complete description of slip planes in plasticity;
	see \cite{Piotr-plasticity}.
}
In both frameworks, we choose rather the Eulerian fields as~unknowns or state variables,
i.e. we will work with $(\rho,\vv,\FFF\kpt,\fibC)$.

\subsection{Natural Configurations Framework}
Let us now develop the coupling between the vector field $\fibC$ describing the anisotropy
and motion of the elastic material within NCF.
Since the balance laws and thermodynamics developed in~\ref{subsubsec:BalanceLaws} and \ref{subsubsec:NCFThermodynamics},
respectively, remain valid, we may move directly to~enhancing the kinematics of NCF.

\subsubsection{Kinematics of the Vector Describing Anisotropy}
Building on the concept of~the~natural configuration described in~\ref{subsubsection:NCFKinematics},
we see that the~relation between the vector $\fibR$ in~the~reference configuration
and the~vector $\fibC$ in~the current configuration can be rewritten using the decomposition
\eqref{eq:NatConfSplit}
\begin{align*}
	\fibC = \FFF \fibR = \FFF\kpt \GGG \fibR = \FFF\kpt \fibN,
	\quad
	\text{where we defined}
	\quad
	\fibN := \GGG \fibR.
\end{align*}
The material derivative of $\fibC$ is then
\begin{align*}
	\dot{\fibC} = \dot{\FFF}\kpt \fibN + \FFF\kpt \dotfibN.
\end{align*}

One may suppose the vector $\fibN$, describing the~material symmetry in~the~natural configuration,
has a purely reversible evolution, i.e. $\dotfibN = 0$.
This tells
us
that the material symmetry is not affected by the dissipative deformation $\GGG$
and remains fixed in the~natural configuration.
This is common for instance for lattice vectors in crystal plasticity;
see e.g. \cite{GurtinAnand2010,Piotr-plasticity}.

\subsubsection{Derivation of Models}
The derivation is again based on~prescribing two scalar quantities,
the~dissipation rate $\xi$ and the internal Helmholtz free energy.
Here we assume
\begin{equation*}
	\psi=\psi(T,\rho,\F\kpt,\fibC)
    =
    \rho\psi_0(T, \rho)
    +
	\rho \psi_\el(T, \FFF\kpt,\fibC).
\end{equation*}
The reduced thermodynamic identity \eqref{reduced_thermodynamic_identity} then becomes,
using the symmetry of $\displaystyle\pder{\psi}{\Fp}\Fp^\top$ and $\displaystyle\pder{\psi}{\fibC} \otimes \fibC$,
\begin{align}\label{reduced_TDI}
	\xi
	=&
	\TTT : \LLL - \rho \dot \psi
	=
	\TTT : \LLL - \rho\pder{\psi_0}{\rho}\dot{\rho} - \rho\pder{\psi_\el}{\FFF\kpt} \cdot \dot{\FFF}\kpt - \rho\pder{\psi_\el}{\fibC} \cdot \dot{\fibC} \\
	=&
	\left(\TTT + p_{\text{th}} \III - \rho\pder{\psi_\el}{\Fp}\Fp^\top - \rho\pder{\psi_\el}{\fibC} \otimes \fibC \right) : \DDD
		+ \rho\pder{\psi_\el}{\Fp} : (\LLL\Fp - \dot{\F}\kpt) + \rho\pder{\psi_\el}{\fibC} \cdot ( \LLL \fibC - \dot{\fibC}) \nonumber\\
	=&
	\left(\TTT + p_{\text{th}} \III - \rho\pder{\psi_\el}{\Fp}\Fp^\top - \rho\pder{\psi_\el}{\fibC} \otimes \fibC \right) : \DDD
	+ \left( \rho\pder{\psi_\el}{\Fp}\Fp^\top + \rho\pder{\psi_\el}{\fibC} \otimes \fibC \right) : (\LLL - \dot{\F}\kpt\Fp^{-1})\nonumber\\
		&+ \rho\pder{\psi_\el}{\fibC} \cdot ( \dot{\F}\kpt\Fp^{-1} \fibC - \dot{\fibC}),\nonumber
\end{align}
where again $p_{\text{th}}$ is the~hydrodynamic pressure defined in~\eqref{eq.p.psi}.
The last form of~the~dissipation rate is suitable for~closing the~system when
$\dotfibN = 0$, i.e. 
\begin{equation}
\dot{\fibC} = \dot{\FFF}\kpt \fibN = \dot{\FFF}\kpt \Fp^{-1} \fibC,\label{dotfibC}
\end{equation}
or when the~material dissipates in the total elastic Cauchy stress
\begin{align*}
	\rho\pder{\psi_\el}{\Fp}\Fp^\top + \rho\pder{\psi_\el}{\fibC} \otimes \fibC.
\end{align*}

\subsubsection{Specific Example}
To~enhance the isotropic Giesekus model from the previous chapter
we modify the prescribed dissipation
\begin{equation}\label{diss001}
	\xi
	=
	2\mu|\D|^2 + \Lambda \left| \rho\pder{\psi_\el}{\Fp}\Fp^\top + \rho\pder{\psi_\el}{\fibC} \otimes \fibC \right|^2,
\end{equation}
where the second term,
containing the total elastic Cauchy stress,
corresponds to a~dissipation caused by the evolution of~the~natural configuration
in~which the~anisotropy remains constant.
Upon comparing \eqref{reduced_TDI} and \eqref{diss001}, one gets
\begin{subequations}\label{eq001}
\begin{align}
	\TTT
	&=
	-p_{\text{th}}\III + 2\mu\D + \Tel,\\
	\Tel
	&=
	\rho\pder{\psi_\el}{\Fp}\Fp^\top + \rho\pder{\psi_\el}{\fibC} \otimes \fibC,\label{eq001b}\\
	\dot{\F}\kpt
	&=
	\LLL \Fp - \Lambda \Tel \Fp,\label{eq001c} \\
	\dot{\fibC} &= \LLL \fibC - \Lambda \Tel \fibC,
\end{align}
\end{subequations}
where in the last equation we combined \eqref{eq001c} and \eqref{dotfibC}.

To close \eqref{eq001b} we assume a special form of the internal free energy
\begin{align*}
	\psi_\el(T, \FFF\kpt,\fibC)
    &=
    \frac{G(T)}{2} (|\FFF\kpt|^2 - 3)
    -
    k_B T \ln \det (\FFF\kpt \FFF\kpt^\top)
    +
    \frac{H(T)}{2} \left( \frac{|\fibC|^2}{L^2(T)} - 1 \right)
    -
    k_B T \ln \frac{|\fibC|^2}{L^2(T)},
\end{align*}
i.e. an~anisotropic modification of~the compressible neo-Hookean solid free energy.
The material constant $L(T)$ is introduced for dimensional reasons and represents the~vector's stress-free length.
The partial derivatives of~$\psi_\el$ being
\begin{align}\label{psiderivative}
	\pder{\psi_\el}{\Fp}
	&=
	G(T) \left( \Fp - \frac{2k_B T}{G(T)}\Fp^{-\top} \right), \\
	\pder{\psi_\el}{\fibC}
	&=
	H(T) \left( \frac{|\fibC|^2}{L^2(T)} - \frac{2 k_B T}{H(T)} \right) \frac{\fibC}{|\fibC|^2},
\end{align}
the equation \eqref{eq001b} becomes
\begin{align*}
	\Tel
	&=
	\rho G(T) \left( \Fp \Fp^\top - \frac{2k_B T}{G(T)} \III \right)
	+ \rho H(T) \left( \frac{|\fibC|^2}{L^2(T)} - \frac{2 k_B T}{H(T)} \right) \frac{\fibC \otimes \fibC}{|\fibC|^2}.
\end{align*}
As~for ~the isotropic model we obtain the complete set of evolution equations for $(\rho, \vv, \BBB\kpt, \fibC)$
\begin{subequations}
\label{eq:NCFFinal}
\begin{align}
	\dot{\rho} &= -\rho \dive \vv, \\
	\rho \dot{\vv} &= \dive ( - p_{\rm th}\I + 2\mu\D + \Tel ), \\
	\dot{\BBB}\kpt
	&=
	\LLL\BBB\kpt + \BBB\kpt \LLL^\top - \Lambda \BBB\kpt \Tel - \Lambda \Tel \BBB\kpt,\\
	\dot{\fibC} &= \LLL \fibC - \Lambda \Tel \fibC, \\
	\nonumber
	\Tel
	&=
	\rho G(T) \left( \BBB\kpt - \frac{2k_B T}{G(T)} \III \right)
		+ \rho H(T) \left( \frac{|\fibC|^2}{L^2(T)} - \frac{2 k_B T}{H(T)} \right) \frac{\fibC \otimes \fibC}{|\fibC|^2},
\end{align}
\end{subequations}
a variant of an~anisotropic Giesekus model using only the fully Eulerian fields,
identical to equations obtained by~GENERIC in~\eqref{eq.gen.aniso.final}.

\begin{remark}
	As~in~the~isotropic case, an~anisotropic modification of~the Oldroyd-B
	model is determined by
	\begin{equation*} 
		\xi
		=
		2\mu|\D|^2+\Lambda\left|\rho\pder{\psi_\el}{\Fp}+\rho\pder{\psi_\el}{\fibC}\otimes\Fp^{-1}\fibC\right|^2,
	\end{equation*}
	depending rather
	on~the stress tensor between the~natural and the~current configuration.
	Anisotropic Maxwell model is obtained for $\mu=0$.
	Additional anisotropic viscosity may be achieved by adding the term $2 \nu \dfrac{\left( \D\fibC \cdot \fibC \right)^2}{|\fibC|^4}$.
\end{remark}

In summary, we have used the NCF to derive models involving anisotropy characterized by the vector field $\fibC$.
When analogical internal free energy and dissipation potential are invoked within GENERIC, the same models can be obtained,
as we will see in the next subsection.

\subsection{GENERIC}

The procedure of deriving the field equations is the same as for the isotropic model,
only more information from the Lagrangian description by $(\yy,\MM)$ is extracted.
The~derivation of the mechanics for $(\rho,\mm,\AAA,\fibC)$
relies completely on our understanding of the nature of $\fibC$.
We suppose that the anisotropy in the reference configuration
is described by a \emph{vector} field $\fibR(\XX)$
and hence we define
\begin{align} \label{eq:FiberDef}
	\left. \fibC(\xx) \right|_{\xx = \yy(\XX)}
	:=
	\frac{\partial \yy}{\partial \XX}(\XX) \BB(\XX).
\end{align}
Although strictly speaking, the anisotropy
is rather described by the structural tensors $\fibC \otimes \fibC$
and $\fibR \otimes \fibR$
(they do not carry the information about orientation,
only about the one-dimensional subspace,
i.e. they are invariant under changing the direction of the vector),
for simplicity we rather choose $\fibC$ as the additional state variable.

\subsubsection{Reversible Kinematics}
As already mentioned in~the precedent section \ref{sec:iso},
the projection of $(\yy,\MM)$ on $(\rho,\mm,\AAA)$
is carried out in~\cite[App. B]{lageul}.
Here we therefore highlight just how the new field $\fibC$ enters the procedure.
We hence consider a~projection from $(\yy, \MM)$
to $(\rho, \mm, \AAA, \fibC)$.
Recalling the definition \eqref{eq:FiberDef}
we have
\begin{align*}
  \fibCi^i(\xx)
  :=
  F^i_I (\xx) \fibRi^I(\xx),
\end{align*}
where $F^i_I$ stands for the deformation gradient
(i.e. the inverse of the distortion $A^J_j$) and
\begin{align*}
  \left. \fibR (\xx) \right|_{\xx = \yy(\XX)}
  :=
  \fibR(\XX)
\end{align*}
is the Eulerian field derived
from the Lagrangian field $\fibR(\XX)$. The $\fibC$  vector can be interpreted also as the orientation vector from \cite{Beris-anisotropic}, where its evolution is equipped also with inertial terms stemming from rigid body rotations.

Proceeding as~in~\cite{lageul} we arrive at~a~new Poisson bracket
\begin{equation}
	\label{eq.PB.Bb}
	\begin{split}
		\{F,G\}^{\text{A\&\fibCi}}
		&=
		\{F,G\}^{\text{A}} \\
	  & \quad
	  +
	  \int_{\RRR^3}
		\fibCi^i
		\left(
		    \frac{\delta F}{\delta \fibCi^j}
		    \frac{\partial}{\partial x^i}
		    \frac{\delta G}{\delta m_j}
		    -
		    \frac{\delta G}{\delta \fibCi^j}
		    \frac{\partial}{\partial x^i}
		    \frac{\delta F}{\delta m_j}
		\right) \,
	  \d \xx \\
	  & \quad
	   -
	   \int_{\RRR^3}
		 \frac{\partial \fibCi^j}{\partial x^i}
		 \left(
		     \frac{\delta F}{\delta \fibCi^j}
		     \frac{\delta G}{\delta m_i}
		     -
		     \frac{\delta G}{\delta \fibCi^j}
		     \frac{\delta F}{\delta m_i}
		 \right) \,
		\d \xx,
	\end{split}
\end{equation}
where bracket $\{F,G\}^{A}$ was defined in \eqref{eq.PB.Bb};
for more detailed computation see Appendix \ref{sec.PB.Bb}.
This bracket (when dropping the distortion field) is similar a part of a Poisson bracket from \cite{Edwards-liquid,Beris-anisotropic},
but in~our work the vector field $\fibC$ is advected (or Lie-dragged) as a~proper vector field rather than
as a~triplet of~scalar fields.

The standard localization procedure, the same used for isotropic material,
leads to field equations
\begin{subequations}
\label{eq:AniRev}
\begin{align}
    \left( \frac{\partial \rho}{\partial t} \right)_{\text{rev}}
    &=
    - \frac{\partial}{\partial x^i} \left( \rho \frac{\delta E}{\delta m_i} \right), \\
    \left( \frac{\partial m_i}{\partial t} \right)_{\text{rev}}
    &=
    - \frac{\partial}{\partial x^j} \left( m_i \frac{\delta E}{\delta m_j} \right)
    - \rho \frac{\partial}{\partial x^i} \frac{\delta E}{\delta \rho}
    - m_j \frac{\partial}{\partial x^i} \frac{\delta E}{\delta m_j}\\
    &\quad - A^J_j \frac{\partial}{\partial x^i} \frac{\delta E}{\delta A^J_j}
     + \frac{\partial}{\partial x^i}
    \left( A^J_j \frac{\delta E}{\delta A^J_j}
            + \frac{\delta E}{\delta \fibCi^j} \fibCi^j \right)\nonumber \\
    & \quad
    - \fibCi^j \frac{\partial}{\partial x^i} \frac{\delta E}{\delta \fibCi^j} 
    + \frac{\partial}{\partial x^j}
    \left( -A^J_i \frac{\delta E}{\delta A^J_j}
            + \frac{\delta E}{\delta \fibCi^i} \fibCi^j \right)\nonumber \\
    \left( \frac{\partial A^I_i}{\partial t} \right)_{\text{rev}}
    &=
    - \frac{\partial A^I_i}{\partial x^j} \frac{\delta E}{\delta m_j}
    -A^I_j \frac{\partial}{\partial x^i} \frac{\delta E}{\delta m_j}, \\
    \left( \frac{\partial \fibCi^i}{\partial t} \right)_{\text{rev}}
    &=
    - \frac{\partial \fibCi^i}{\partial x^j} \frac{\delta E}{\delta m_j}
    + \frac{\partial}{\partial x^j} \frac{\delta E}{\delta m_i} \fibCi^j.
\end{align}
\end{subequations}
Since the evolution equation for $\fibC$ is also compatible
with the equation $\dot{\FFF} = \LLL \FFF$,
it is again the equation of momentum which
is not in the standard form.
Nevertheless, it is also valid for any energy functional.
In~order to highlight the resemblance with the other models as before,
we stick in the next section to~an energy ansatz,
similar to the one for the isotropic model.

\subsubsection{Elastic Dynamics}
As~in the~preceding section we stick to~isothermal processes,
using a~modification of~the free energy ansatz \eqref{eq:Helmholtz}
\begin{align} \label{eq:HelmholtzAni}
	\nonumber
    \tilde{F}(\rho, \mm, \AAA, \fibC)
    &=
    F(\rho, \mm, T, \AAA, \fibC)
    =
    \int_{\RRR^3}
        f(\rho, \mm, T, \AAA, \fibC) \,
    \d \xx \\
    &=
    \int_{\RRR^3}
        \rho \psi(\rho, \mm, T, \AAA, \fibC) \,
    \d \xx
    =
    \int_{\RRR^3}
        \frac{|\mm|^2}{2 \rho}
        +
        \rho \psi_0(T, \rho)
        +
        \rho \psi_\el(T, \AAA, \fibC) \,
    \d \xx,
\end{align}
for which the equation for~momentum simplifies to
\begin{align*}
    \left( \frac{\partial m_i}{\partial t} \right)_{\text{rev}}
    &=
    - \frac{\partial}{\partial x^j} \left( m_i \frac{m_j}{\rho} \right)
    - \rho \frac{\partial}{\partial x^i} \frac{\partial f}{\partial \rho}
    - m_j \frac{\partial}{\partial x^i} \frac{\partial f}{\partial m_j}
    \\
    & \quad
    - A^J_j \frac{\partial}{\partial x^i} \frac{\partial f}{\partial A^J_j}
    - \fibCi^j \frac{\partial}{\partial x^i} \frac{\partial f}{\partial \fibCi^j}
    + \frac{\partial}{\partial x^i}
    \left(
        A^J_j \frac{\partial f}{\partial A^J_j}
        + \frac{\partial f}{\partial \fibCi^j} \fibCi^j
    \right)
    \frac{\partial}{\partial x^j}
    \left(
        - A^J_i \frac{\partial f}{\partial A^J_j}
        + \frac{\partial f}{\partial \fibCi^i} \fibCi^j
    \right) \\
    &
    =
    - \frac{\partial}{\partial x^j} \left( m_i \frac{m_j}{\rho} \right)
    - \frac{\partial}{\partial x^i}
    \left(
        - f
        + \rho \frac{\partial f}{\partial \rho}
        + m_j \frac{\partial f}{\partial m_j}
    \right)
	+ \frac{\partial}{\partial x^j}
    \left(
        - A^J_i \frac{\partial f}{\partial A^J_j}
        + \frac{\partial f}{\partial \fibCi^i} \fibCi^j
    \right) \\
    &
    =
    - \frac{\partial}{\partial x^j} \left( m_i \frac{m_j}{\rho} \right)
    - \frac{\partial}{\partial x^i}
    \left(
        \rho^2 \frac{\partial \psi_0}{\partial \rho}
    \right)
    + \frac{\partial}{\partial x^j}
    \left(
        - \rho A^J_i \frac{\partial \psi}{\partial A^J_j}
        + \rho \frac{\partial \psi}{\partial \fibCi^i} \fibCi^j
    \right).
\end{align*}
Hence we arrive at the system
\begin{align*}
    \left( \frac{\partial \rho}{\partial t} \right)_{\text{rev}}
    &=
    - \dive \left( \rho \vv \right), \\
   \left( \frac{\partial \mm}{\partial t} \right)_{\text{rev}}
    &=
    - \dive \left( \mm \otimes \vv \right)
    + \dive (-p_{th}\III + \Tel), \\
    \left( \frac{\partial \AAA}{\partial t} \right)_{\text{rev}}
    &=
    - (\nabla \AAA) \vv
    - \AAA \LLL, \\
    \left( \frac{\partial \fibC}{\partial t} \right)_{\text{rev}}
    &=
    - [\nabla \fibC] \vv
    + \LLL \fibC, \\
    \Tel
    &=
    - \rho \AAA^{\top} \frac{\partial \psi_\el}{\partial \AAA}
    + \rho \frac{\partial \psi_\el}{\partial \fibC} \otimes \fibC,
\end{align*}
where again we use the same notation as for the isotropic material
\begin{align*}
    \vv := \frac{\delta F}{\delta \mm}
    = \frac{\partial f}{\partial \mm} = \frac{\mm}{\rho},
    \qquad
    p_{th} := \rho^2 \frac{\partial \psi_0}{\partial \rho},
\end{align*}
the only difference being in the~elastic part of the~Cauchy stress $\Tel$,
which now contains the~anisotropic elastic response.
Rewriting the~system in~terms of~$\Fp := \AAA^{-1}$ with the~convective derivative yields
\begin{subequations}
\begin{align*}
    \dot{\rho} &= -\rho \dive \vv, \\
    \rho \dot{\vv}
    &=
    \dive
    \left(
        -p_{th} \III + \Tel
    \right), \\
    \dot{\FFF}\kpt
    &=
    \LLL \FFF_\kappa, \\
    \dot{\fibC}
    &=
    \LLL \fibC, \\
    \Tel
    &=
    \rho \frac{\partial \psi_\el}{\partial \FFF_\kappa}\FFF_\kappa^\top
    	+ \rho \frac{\partial \psi_\el}{\partial \fibC} \otimes \fibC.
\end{align*}
\end{subequations}

If~one were to describe an anisotropic material,
where $\fibC$ describes the orientation of elastic fibers,
a~suitable candidate for $\psi_\el$, based on the former one in \eqref{eq:HelmholtzA},
would be
\begin{equation}
	\label{eq:HelmholtzAniFib}
	\begin{split}
		\psi_\el(T,\AAA,\fibC)
		&=
		\frac{G(T)}{2} (|\AAA^{-1}|^2 - 3)
		-
		k_B T \ln \det (\AAA^{-1} \AAA^{-\top})
		+ \frac{H(T)}{2} \left( \frac{|\fibC|^2}{L^2(T)} - 1 \right)
		- k_B T \ln \frac{|\fibC|^2}{L^2(T)},
    \end{split}
\end{equation}
or equivalently in~$\Fp$
\begin{equation}
	\label{eq:HelmholtzAniFibF}
	\begin{split}
		\psi_\el(T,\Fp,\fibC)
		&=
		\frac{G(T)}{2} (|\Fp|^2 - 3)
		-
		k_B T \ln \det (\Fp \Fp^{\top})
		+ \frac{H(T)}{2} \left( \frac{|\fibC|^2}{L^2(T)} - 1 \right)
		- k_B T \ln \frac{|\fibC|^2}{L^2(T)},
    \end{split}
\end{equation}
where the material constant $L(T)$ is introduced for~dimensional reasons
and represents the fiber's stress-free length.
Indeed, for this particular $\psi_\el$ we have
\begin{align}
	\label{eq:ElCauchyAniA}
    \Tel
    &=
    \rho G(T) \left( \AAA^{-1}\AAA^{-\top} - \frac{2 k_B T}{G(T)} \III \right)
    + \rho H(T) \left( \frac{|\fibC|^2}{L^2(T)} - \frac{2 k_B T}{H(T)}\right)
    	\frac{\fibC \otimes \fibC}{|\fibC^2|} \\
    &=
    \label{eq:ElCauchyAniF}
    \rho G(T) \left( \Fp\Fp^{\top} - \frac{2 k_B T}{G(T)} \III \right)
    + \rho H(T) \left( \frac{|\fibC|^2}{L^2(T)} - \frac{2 k_B T}{H(T)}\right)
    	\frac{\fibC \otimes \fibC}{|\fibC^2|}.
\end{align}

\subsubsection{Irreversible Kinematics}
As~in the previous section, we will start to~design the~isothermal irreversible evolution
by~using the~corresponding part of~the~equation \eqref{eq.gen.iso}.
Now we have $\tilde{\qq} = (\rho,\mm,\AAA)$ and $\Xi = \Xi(\AAA,\fibC,\AAA^*,\fibC^*)$ quadratic as in~\eqref{eq:DissPotQuad}.
We hence obtain
\begin{equation}
	\label{eq:BbIrr}
	\begin{split}
		\left( \frac{\partial A^I_i}{\partial t} \right)_{\text{irr}}
		&=
		-\frac{1}{T}
		\left.
		    \frac{\delta \Xi}{\delta (A^*)^i_I}
		\right|_{(A^*)^i_I = \frac{\delta F}{\delta A^I_i}}, \\
		\left( \frac{\partial \fibCi^i}{\partial t} \right)_{\text{irr}}
		&=
		-\frac{1}{T}
		\left.
		    \frac{\delta \Xi}{\delta (\fibCi^*)_i}
		\right|_{(\fibCi^*)_i = \frac{\delta F}{\delta \fibCi^i}}.
	\end{split}
\end{equation}
For $\tilde{\qq} = (\rho,\mm,\Fp)$ and $\Xi = \Xi(\Fp,\fibC,\Fp^*,\fibC^*)$ the first equation again transforms to~\eqref{eq:Firr}.

\subsubsection{Dissipative Dynamics}

Let us now move to deriving an~anisotropic variant
of an~isothermal Giesekus viscoelastic model.
Using the~total Helmholtz free energy ansatz from \eqref{eq:HelmholtzAni},
the conjugate variables are then replaced by
\begin{align*}
    \mm^* &\rightarrow
    \frac{\partial f}{\partial \mm} = \frac{\mm}{\rho}, \\
    \AAA^* &\rightarrow
    \frac{\partial f}{\partial \AAA} = \rho \frac{\partial \psi_\el}{\partial \AAA}, \\
    \Fp^* &\rightarrow
    \frac{\partial f}{\partial \Fp} = \rho \frac{\partial \psi_\el}{\partial \Fp}, \\
    \fibC^* &\rightarrow
    \frac{\partial f}{\partial \fibC} = \rho \frac{\partial \psi_\el}{\partial \fibC}.
\end{align*}
Our choice of~dissipation potential is
\begin{align}
	\nonumber
    \Xi(\AAA,\fibC,\mm^*,\AAA^*,\fibC^*)
    \nonumber
    &=
    T
    \int_{\RRR^3}
        \left(\mu \left|(\nabla \mm^*)_{sym} \right|^2
        +
        \frac{\Lambda}{2} \left| -\AAA^\top \AAA^* + \fibC^* \otimes \fibC \right|^2\right) \,
    \d \xx \\
	\label{eq:DissPotAniA}
    &=
    T
    \int_{\RRR^3}
        \left(\mu \left|\DDD \right|^2
        +
        \frac{\Lambda}{2} \left| -\rho \AAA^\top \frac{\partial \psi_\el}{\partial \AAA} + \rho \pder{\psi_\el}{\fibC} \otimes \fibC \right|^2\right) \,
    \d \xx,
\end{align}
or equivalently
\begin{align}
	\nonumber
    \Xi(\Fp,\fibC,\mm^*,\Fp^*,\fibC^*)
    &=
    T
    \int_{\RRR^3}
        \left(\mu \left|(\nabla \mm^*)_{sym} \right|^2
        +
        \frac{\Lambda}{2} \left| \Fp^* \Fp^\top + \fibC^* \otimes \fibC \right|^2\right) \,
    \d \xx \\
	\nonumber
    &\quad=
    T \int_{\RRR^3}
        \left(\mu \left|\DDD \right|^2
        +
        \frac{\Lambda}{2} \left| \rho \frac{\partial \psi_\el}{\partial \Fp} \Fp^\top + \rho \pder{\psi_\el}{\fibC} \otimes \fibC \right|^2\right) \,
    \d \xx
	\label{eq:DissPotAniF}
    =
    \hat{\Xi}(\Fp,\fibC,\Fp^* \Fp^\top + \fibC^* \otimes \fibC),
\end{align}
where again $\DDD$ stands for the symmetric velocity gradient,
the first term yields the standard viscous dissipation,
and the second is chosen as being proportional to the elastic Cauchy stress,
now with an~anisotropic contribution.
As~we will see later,
this choice of~dissipation potential
leaves the lattice vectors in the~natural configuration constant,
i.e. they are invariant under the plastic deformation;
see e.g. \cite[Chap. 91]{GurtinAnand2010}.
This is not an~coincidence;
it can be shown by~convex analysis that the dissipation potential
depends on~the~conjugate variables in~this special way,
i.e. $\Xi(\Fp,\fibC,\mm^*,\Fp^*,\fibC^*) = \hat{\Xi}(\Fp,\fibC,\Fp^* \Fp^\top + \fibC^* \otimes \fibC)$,
if and only if the~irreversible evolution written in~terms of~the~dual dissipation potential $\Xi^*$
satisfies the~constraint $\dot{\fibC} = \dot{\FFF}\kpt \Fp^{-1} \fibC$.
Here the vector $\fibC$ corresponds to the observed lattice vector,
i.e. mapped from the~natural configuration by $\Fp = \AAA^{-1}$.
Computing the derivatives of $\Xi$
\begin{align*}
    \frac{\delta \Xi}{\delta \mm^*}
    &=
    -T \dive(2 \mu  (\nabla \mm^*)_{sym})
    = - T \dive(2 \mu \DDD), \\
    \frac{\delta \Xi}{\delta \AAA^*}
    &=
    - T \Lambda \AAA \left( -\AAA^\top \AAA^* + \fibC^* \otimes \fibC \right)
    =
    -T \Lambda \AAA \left( - \rho \AAA^\top \frac{\partial \psi_\el}{\partial \AAA}  + \rho \pder{\psi_\el}{\fibC} \otimes \fibC\right), \\
    \frac{\delta \Xi}{\delta \Fp^*}
    &=
    T \Lambda \left( \Fp^* \Fp^\top + \fibC^* \otimes \fibC \right) \Fp
    =
    T \Lambda \left( \rho \frac{\partial \psi_\el}{\partial \Fp} \Fp^\top + \rho \pder{\psi_\el}{\fibC} \otimes \fibC \right) \Fp, \\
    \frac{\delta \Xi}{\delta \fibC^*}
    &=
    T \Lambda \left( \Fp^* \Fp^\top + \fibC^* \otimes \fibC \right) \fibC
    =
    T \Lambda \left( \rho \frac{\partial \psi_\el}{\partial \Fp} \Fp^\top + \rho \pder{\psi_\el}{\fibC} \otimes \fibC \right) \fibC,
\end{align*}
yields the irreversible evolution
\begin{align*}
    \left(\frac{\partial \mm}{\partial t}\right)_{\irr}
    &=
    \dive(2 \mu \DDD), \\
    \left(\frac{\partial \AAA}{\partial t}\right)_{\irr}
    &=
    \Lambda \AAA \Tel, \\
    \left(\frac{\partial \Fp}{\partial t}\right)_{\irr}
    &=
    - \Lambda \Tel \Fp, \\
    \left( \frac{\partial \fibC}{\partial t} \right)_{\text{irr}}
    &=
    - \Lambda \Tel \fibC, \\
    \Tel
    &=
    -\rho \AAA^{\top} \frac{\partial \psi_\el}{\partial \AAA}
    +
    \rho \frac{\partial \psi_\el}{\partial \fibC}
    \otimes
    \fibC
    =
    \rho \frac{\partial \psi_\el}{\partial \Fp} \Fp^\top
    +
    \rho \frac{\partial \psi_\el}{\partial \fibC}
    \otimes
    \fibC.
\end{align*}
In particular, choosing $\psi_\el$ as in \eqref{eq:HelmholtzAniFib} and \eqref{eq:HelmholtzAniFibF}
yields $\Tel$ from \eqref{eq:ElCauchyAniA} and \eqref{eq:ElCauchyAniF}, respectively.

\subsection*{Final Equations}

Let the state variables be represented by $(\rho,\mm,\AAA,\fibC)$.
Their reversible evolution has been specified by the Poisson bracket \eqref{eq.PB.Bb},
leading to equations \eqref{eq:AniRev}.
Once the total free energy is specified, as in \eqref{eq:HelmholtzAni},
the reversible evolution can be written down explicitly.
The irreversible evolution is given by dissipation potential,
e.g. the one in \eqref{eq:DissPotAniA},
and becomes explicit when the concrete formula for total free energy (or entropy in the non-isothermal case) is invoked.
The final evolution equations are then the sum of the reversible and irreversible parts of the evolution
\begin{subequations}
\label{kinnatani}
\begin{align}
    \frac{\partial \rho}{\partial t} + \dive (\rho \vv) &= 0, \\
    \frac{\partial \mm}{\partial t} + \dive(\mm \otimes \vv)
    &=
    \dive
    \left(
        - p_{th} \III + 2 \mu \DDD + \Tel
    \right), \\
    \frac{\partial \AAA}{\partial t} + (\nabla \AAA) \vv
    &=
    - \AAA \LLL + \Lambda \AAA \Tel, \\
    \frac{\partial \fibC}{\partial t} + [\nabla \fibC] \vv
    &=
    \LLL \fibC - \Lambda \Tel \fibC, \\
    \Tel
    &=
    -\rho \AAA^{\top} \frac{\partial \psi_\el}{\partial \AAA}
    +
    \rho \frac{\partial \psi_\el}{\partial \fibC}
    \otimes
    \fibC.
\end{align}
\end{subequations}
Using the~convective time derivative and $\FFF_\kappa := \AAA^{-1}$ instead of the distortion, they can be rewritten to
\begin{subequations}
\begin{align}
    \dot{\rho} &= -\rho \dive \vv, \\
    \rho \dot{\vv}
    &=
    \dive
    \left(
        -p_{th} \III + 2 \mu \DDD + \Tel
    \right), \\
    \dot{\FFF}\kpt
    &=
    \LLL \FFF_\kappa - \Lambda \Tel \FFF_\kappa,\label{eq82c} \\
    \dot{\fibC}
    &=
    \LLL \fibC - \Lambda \Tel \fibC, \\
    \Tel
    &=
    \rho \frac{\partial \psi_\el}{\partial \Fp} \Fp^\top
    +
    \rho \frac{\partial \psi_\el}{\partial \fibC}
    \otimes
    \fibC\label{eq82e},
\end{align}
\end{subequations}
and in terms of~the left Cauchy--Green tensor $\BBB := \AAA^{-1} \AAA^{-\top} = \FFF_\kappa \FFF_\kappa^\top$
the equations \eqref{eq82c} and \eqref{eq82e} become
\begin{subequations}
\begin{align*}
    \dot{\BBB}\kpt
    &=
    \LLL \BBB_\kappa
    +
    \BBB_\kappa \LLL^\top
    - \Lambda \Tel \BBB_\kappa - \Lambda \BBB_\kappa \Tel, \\
    \Tel
    &=
    2 \rho \frac{\partial \psi_\el}{\partial \BBB_\kappa}\BBB_\kappa
    +
    \rho \frac{\partial \psi_\el}{\partial \fibC} \otimes \fibC,
\end{align*}
\end{subequations}
where we can easily recognize the upper-convective derivative in the equation for $\BBB$.
This is compatible with the literature\cite{Miroslav-Ericksen}, where the same result is derived from the two-particle kinetic theory.
If we use the concrete energy \eqref{eq:HelmholtzA},
now written in terms of $\BBB_\kappa$ as
\begin{align*}
    \psi_\el(T,\BBB_\kappa)
    &=
    \frac{G(T)}{2} (\tr \BBB_\kappa - 3)
    -
    k_B T \ln \det (\BBB_\kappa)
	+
	\frac{H(T)}{2} \left( \frac{|\fibC|^2}{L^2(T)} - 1 \right)
	-
	k_B T \ln \frac{|\fibC|^2}{L^2(T)},
\end{align*}
we recover an~anisotropic modification of~a~variant of the~compressible Giesekus model from the previous section
\begin{subequations}
\label{eq.gen.aniso.final}
\begin{align}
    \dot{\rho} &= -\rho \dive \vv, \\
    \rho \dot{\vv}
    &=
    \dive
    \left(
        -p_{th} \III
        + 2 \mu \DDD
        + \Tel
    \right), \\
    \dot{\BBB}\kpt
    &=
    \LLL \BBB_\kappa
    +
    \BBB_\kappa \LLL^\top
    - \Lambda \Tel \BBB_\kappa - \Lambda \BBB_\kappa \Tel, \\
    \nonumber
    \Tel
    &=
    \rho G(T) \left( \BBB\kpt - \frac{2 k_B T}{G(T)} \III \right)
    + \rho H(T) \left( \frac{|\fibC|^2}{L^2(T)} - \frac{2 k_B T}{H(T)}\right)
    	\frac{\fibC \otimes \fibC}{|\fibC^2|}.
\end{align}
\end{subequations}
These equations describe evolution of~the~state variables in~the~isothermal regime
and are~the~very same as~the~equations obtained within NCF in~\eqref{eq:NCFFinal} 

\begin{remark}
	An~anisotropic modification of~the~Oldroyd-B model would be given by
	\begin{align*}
		\Xi(\AAA,\fibC,\mm^*,\AAA^*,\fibC^*)
		&=
		T \int_{\RRR^3}
		    \left(\mu \left|(\nabla \mm^*)_{sym} \right|^2
		    +
		    \frac{\Lambda}{2} \left| -\AAA^\top \AAA^* \AAA^\top + \fibC^* \otimes \AAA \fibC \right|^2\right) \,
		\d \xx \\
		&=
		T \int_{\RRR^3}
		    \left(\mu \left|\DDD \right|^2
		    +
		    \frac{\Lambda}{2}
		    \left|
		    	-\rho \AAA^\top \frac{\partial \psi_\el}{\partial \AAA} \AAA^\top
		    	+
		    	\rho \frac{\partial \psi_\el}{\partial \fibC} \otimes \AAA \fibC
		    \right|^2\right) \,
		\d \xx, \nonumber \\
		\Xi(\Fp,\fibC,\mm^*,\Fp^*,\fibC^*)
		&=
		T \int_{\RRR^3}
		    \left(\mu \left|(\nabla \mm^*)_{sym} \right|^2
		    +
		    \frac{\Lambda}{2} \left| \Fp^* + \fibC^* \otimes \Fp^{-1} \fibC \right|^2\right) \,
		\d \xx \\
		&=
		T \int_{\RRR^3}
		    \left(\mu \left|\DDD \right|^2
		    +
		    \frac{\Lambda}{2}
		    \left|
		    	\rho \frac{\partial \psi_\el}{\partial \Fp}
		    	+
		    	\rho \frac{\partial \psi_\el}{\partial \fibC} \otimes \Fp^{-1} \fibC
		    \right|^2\right) \,
		\d \xx, \nonumber
	\end{align*}
	i.e. again by changing the~stress measure.
	Dropping the viscous part would yield an~anisotropic modification of~the Maxwell model.
	Additional anisotropic viscosity may be achieved by adding $2 \nu \dfrac{\left( \D\fibC \cdot \fibC \right)^2}{|\fibC|^4}$.
\end{remark}

\section{Conclusion}
In this work we have recalled the framework of natural configurations (NCF) developed by Rajagopal and Srinivasa and the GENERIC framework.
Both approaches are introduced and demonstrated on a variant of the isotropic isothermal Giesekus model, where they are compatible.
We provide a new interpretation of the natural configuration and its evolution within the context of GENERIC.
Conversely, NCF provides an alternative interpretation of the GENERIC for the state variables $(\rho,\mm,\Fp)$, splitting the evolution into the reversible (Hamiltonian) and irreversible (gradient) dynamics. The evolution from the reference configuration to the natural configuration within NCF is the analog of the irreversible part of GENERIC and the evolution from the natural configuration to the current one is the analog of the reversible part of GENERIC.

Both the frameworks are subsequently used to derive an anisotropic model of complex fluids, having an extra vectorial state variable describing the symmetry of the material.
The dissipation can
be
chosen so that both the frameworks again coincide.
We conclude that NCF is compatible with GENERIC.

When studying the two frameworks,
we have also refined the formulation of anisotropic behavior within NCF.
We have also extended the derivation of the Eulerian Poisson brackets from the Lagrangian bracket to include the~vector field $\fibC$ describing anisotropy, using not only the left Cauchy-Green tensor, but the whole distortion.
Finally,
we observed that the~anisotropy in~the~natural configuration
is preserved if and only if the~dissipation potential dependent on~the~elastic Cauchy stress in~the~sense of~\eqref{eq:DissPotAniF}.

The NCF was initially formulated only with the deformation gradient $\Fp$ as the extra state variable beyond fluid mechanics. However, it is easily extended to another state variables evolving together with the continuum (e.g. Lie-dragged in the terminology of Hamiltonian mechanics). Note also that such extra state variables typically do not obey any conservation law, and NCF thus goes beyond the usual Ansatz of non-equilibrium thermodynamic searching for balance laws. Roughly speaking NCF can be extended to all cases where kinematics is constructed from particle mechanics by the Lagrange$\rightarrow$Euler transformation, providing the interpretation of the reversible-irreversible splitting via the natural configuration.
Apart from showing that NCF and GENERIC are compatible, we shed light on the construction of anisotropic continuum thermodynamic models in both frameworks.

\section*{Acknowledgment}
We are indebted to Josef Málek for encouraging this line of research, for teaching us the NCF framework, and for his generous support.
P.P., .K.T., M.P. and M.Š. were supported by Charles University Research program No. UNCE/SCI/023.
P.P., K.T., M.S. and M.Š. were supported by Czech Science Foundation, project no. 18-12719S.
MS was supported by Grant Agency of Charles University, student project no. 282120.



\appendix

\section{Derivation of the entropy for polymeric dumbells}\label{sec.entropy}
Let us now recall how to derive the formula for entropy of a fluid with immersed polymeric dumbells. The derivation is analogical to \cite{Sarti}, but explicitly involves the principle of maximum entropy (MaxEnt), see more in \cite{pkg}. Consider the two-particle Liouville entropy,
\begin{align}\label{eq.entropy.downbasic}
   S^{(L2)} = -k_B \int\int\int \int &
     f(\rr_1,\pp_1,\rr_2,\pp_2)\ln \left(h^6 f(\rr_1,\pp_1,\rr_2,\pp_2)\right)
    \d \rr_1 \d \pp_1 \d \rr_2 \d \pp_2,\nonumber
\end{align}
where $\rr_1, \pp_1, \rr_2, \pp_2$ are the positions and momenta of the two beads at the ends of the dumbells. This entropy can be maximized keeping the constraints 
\begin{subequations}\label{eq.entropy.const}
    \begin{align}
        n(\rr) &= \int \int \int f(\rr,\RR,\pp_1,\pp_2) \, \d \RR \d \pp_1\d \pp_2\\
        c^{ij}(\rr) &= \int \int \int R^i R^j f(\rr,\RR,\pp_1,\pp_2) \, \d \RR \d \pp_1\d \pp_2\\
        \epsilon(\rr) &= \int \int \int \left(\frac{\pp^2_1}{2m} + \frac{\pp^2_2}{2m}\right)  f(\rr,\RR,\pp_1,\pp_2) \, \d \RR \d \pp_1\d \pp_2,
    \end{align}
where $\rr=(\rr_1+\rr_2)/2$, $\RR=\rr_2-\rr_1$ and $m$ is the mass of each of the beads.
\end{subequations}
Maximization of the entropy then reads 
\begin{align}
	\nonumber
    0 &= \frac{\delta}{\delta f}\left(-S^{L2}(f)
    +
    \int \left( n^*(\rr) n(f) + c^*_{ij}(\rr) c^{ij}(f) + \epsilon^*(\rr) \epsilon(f)\right)\d \rr \right),
\end{align}
where the conjugate fields (with the star subscript) play the role of Lagrange multipliers. This equation has the solution 
\begin{equation}
    \tilde{f} = \frac{1}{h^6} e^{-1-\frac{n^*}{k_B}} e^{-\frac{c^*_{ij}R^i R^j}{k_B}} e^{-\frac{\epsilon^*}{k_B}\left(\frac{\pp_1^2}{2m}+\frac{\pp_2^2}{2m}\right)},
\end{equation}
which is the MaxEnt estimate of the distribution function based on the knowledge of constraints \eqref{eq.entropy.const}. By plugging the solution to \eqref{eq.entropy.const} we can now get the MaxEnt values for the state variables
\begin{subequations}\label{eq.statevar.nec}
    \begin{align}
        n(\rr) &= \frac{1}{h^6} e^{-1-\frac{n^*}{k_B}} \left(\frac{2\pi m k_B}{\epsilon^*}\right)^3 \frac{(\pi k_B)^{3/2}}{\sqrt{\sigma_1^* \sigma_2^* \sigma_3^* }}\\
        \epsilon(\rr) &= 3 k_B \frac{n}{\epsilon^*}\\
        c^{ij}(\rr) &= \frac{1}{2} n k_B \left((\cc^*)^{-1}\right)^{ij}.
    \end{align}
\end{subequations}
To integrate $\tilde{f}$, it is convenient to change the coordinates $\RR$ to $\hat{\RR}$ for which 
$c^*_{ij} = \frac{\partial \hat{R}^k}{\partial R^i}\frac{\partial \hat{R}^l}{\partial R^j} \diag(\sigma_1^*,\sigma_2^*,\sigma_3^*)_{kl}$ and $R^i = \frac{\partial R^i}{\partial \hat{R}^k} \hat{R}^k$ because then the expression $c^*_{ij}R^i R^j$ splits the sum of three terms, each dependent only on one component of $\hat{\RR}$. Moreove, since $\cc$ is symmetric, the matrix $\frac{\partial \hat{R}^k}{\partial R^i}$ has unit determinant. At this point we could just invert the relations \eqref{eq.statevar.nec} and plug the results to the original entropy \eqref{eq.entropy.downbasic}. This way we would also get the result \eqref{eq.entropy.final}. However, the proper full way to approach the derivation is to invert the relations using further legendre transforms. To obtain the conjugate entropy we can write
\begin{align}
    S^*(n^*, \cc^*, \epsilon^*)
    =
    -S^{L2}(\tilde{f}(n^*, \cc^*, \epsilon^*))
    +
    \int \left( n^*(\rr) n(\tilde{f};r) + c^*_{ij}(\rr) c^{ij}(\tilde{f};r) + \epsilon^*(\rr) \epsilon(\tilde{f};r)\right)\d \rr.
\end{align}
Explicitly the entropy is obtained using the equation for $n$
\begin{align}
    S^*(n^*, \cc^*, \epsilon^*)
    = - \int k_B n(\rr, \epsilon^*, n^*, \cc^*) \d\rr
    = - \frac{k_B}{h^6} \int e^{-1-\frac{n^*}{k_B}}\left(\frac{2\pi m k_B}{\epsilon^*}\right)^3 \frac{(\pi k_B)^{3/2}}{\sqrt{\sigma_1^* \sigma_2^* \sigma_3^*}} \d\rr
\end{align}
In order to get the entropy in terms of the state variables (not the conjugate ones), we have to carry out a further Legendre transform. This way we can obtain the relations for the conjugate variables $n^*, \cc^*$ and $\epsilon^*$.
\begin{align}\label{eq.entropy.LT}
    0=\frac{\delta}{\delta n^*, \cc^*, \epsilon^*} \Big(-S^*(n^*, \cc^*, \epsilon^*)
     + \int \left( n^*(\rr) n(\rr) + c^*_{ij}(\rr) c^{ij}(\rr) + \epsilon^*(\rr) \epsilon(\rr)\right)\d \rr \Big) ,
\end{align}
When this equations are solved, obtaining inverse relations \eqref{eq.statevar.nec}, we can return to the original entropy in the variables $n, \cc$ and $\epsilon$ by the means of Legendre transform
\begin{align}\label{eq.entropy.dS}
    S(n, \cc, \epsilon) = -S^*(n^*, \cc^*, \epsilon^*)
    + \int \left( n^*(\rr) n(\rr) + c^*_{ij}(\rr) c^{ij}(\rr) + \epsilon^*(\rr) \epsilon(\rr)\right)\d \rr.
\end{align}
Entropy \eqref{eq.entropy.dS} then becomes
\begin{align}\label{eq.entropy.final}
    S(n,\cc,\epsilon)
    &=
    \int k_b n + n^*n + \cc^*\cc + \epsilon^*\epsilon \,\d\rr\\
    &=
    k_B \int n\left(\frac{7}{2}-\frac{11}{2}\ln n + 3\ln \epsilon + \frac{1}{2}\ln \det \cc + \ln\frac{16 \sqrt{2} m^3 \pi^{\frac{9}{2}}}{3^3 h^6} \right)\d\rr.
\end{align}
The dependence of the entropy on the $\ln \det \cc$ term can be thus determined by means of statistical physics.

Helmholtz free energy is then constructed as $F = E - TS$ and this is how it inherits the term as well (although with the temperature prefactor). Temperature can be obtained
\begin{equation}
\frac{1}{T}=\frac{\partial S}{\partial \epsilon} = 3 k_B n \frac{1}{\epsilon}.
\end{equation}
Free energy is then calculated easily as 
\begin{align}\label{eq.freeenergy.final}
    F(T,n,\cc)
    =
    E - TS
    =
	k_B
	\int T n \left(-\frac{1}{2}+\frac{11}{2}\ln n - 3\ln \epsilon - \frac{1}{2}\ln \det \cc - \ln\frac{16 \sqrt{2} m^3 \pi^{\frac{9}{2}}}{3^3 h^6} \right)\d\rr.
\end{align}
This free energy is compatible with the standard result\cite{Sarti}.

\section{Details on the derivation of Poisson bracket \eqref{eq.PB.Bb}}\label{sec.PB.Bb}
The purpose of this appendix is to include details of the derivation of Poisson bracket \eqref{eq.PB.Bb}.

Since by the Leibniz rule we have
\begin{align*}
  \frac{\delta \fibCi^i(\xx)}{\delta y^k(\XX)}
  =
  \frac{\delta F^i_I(\xx)}{\delta y^k(\XX)}
  \fibRi^I(\xx)
  +
  F^i_I(\xx) \frac{\delta \fibRi^I(\xx)}{\delta y^k(\XX)},
\end{align*}
we can use the formulas from \cite[App. B]{lageul}
expressing the derivative with respect to arbitrary
Eulerian field
\begin{align*}
  \frac{\delta \fibRi^I(\xx)}{\delta y^k(\XX)}
  =
  - \frac{\partial \fibRi^I}{\partial x^k}(\xx) \delta(\XX-\yy^{-1}(\xx)),
\end{align*}
where $\delta$ is formally treated as the Dirac distribution (see \cite{HCO} for an explanation of this calculus)
and the derivative of the Eulerian field
of the deformation gradient
\begin{align*}
  \frac{\delta F^i_I(\xx)}{\delta y^k(\XX)}
  =
  - \frac{\partial \delta}
         {\partial X^I}(\XX-\yy^{-1}(\xx))
  \delta^i_k
  - \frac{\partial F^i_I}{\partial x^k}(\xx) \delta(\XX-\yy^{-1}(\xx)),
\end{align*}
where $\delta^i_k$ denotes the Kronecker $\delta$.
Hence, we obtain
\begin{align}
  \frac{\delta \fibCi^i(\xx)}{\delta y^k(\xx)}
  &=
  - \frac{\partial \delta}
         {\partial X^I}(\XX-\yy^{-1}(\xx))
  \delta^i_k \fibRi^I(\xx)
  - \frac{\partial F^i_I}{\partial x^k}(\xx) \fibRi^I(\xx) \delta(\XX-\yy^{-1}(\xx))
  - F^i_I(\xx) \frac{\partial \fibRi^I}{\partial x^k}(\xx) \delta(\XX-\yy^{-1}(\xx)) \nonumber \\
  \label{eq:DerrFibCy}
  &=
  - \frac{\partial \delta}
         {\partial X^I}(\XX-\yy^{-1}(\xx))
  \delta^i_k \fibRi^I(\xx)
  - \frac{\partial \fibCi^i}{\partial x^k}(\xx) \delta(\XX-\yy^{-1}(\xx)).
\end{align}

Having in mind the procedure from \cite{lageul}, we can proceed with rewriting of the Poisson bracket
\begin{align*}
  \{B,C\}^{\text{Lagrange}}
  =
  \int_{\RRR^3}
    \left(\frac{\delta B}{\delta y^k(\XX)}
    \frac{\delta C}{\delta M_k(\XX)}
    -
    \frac{\delta C}{\delta y^k(\XX)}
    \frac{\delta B}{\delta M_k(\XX)}\right)\,
  \d \XX.
\end{align*}
Again by \cite[App. B]{lageul} we have
\begin{align}
	\label{eq:DerrGM}
  \frac{\delta C}{\delta M_k(\XX)}
  =
  \left.
    \frac{\delta C}{\delta m_k(\xx)}
  \right|_{\XX = \yy^{-1}(\xx)},
\end{align}
while by the chain rule
\begin{align}
	\label{eq:DerrFy}
  \frac{\delta B}{\delta y^k(\XX)}
  =
  \int_{\RRR^3}
    \frac{\delta B}{\delta \fibCi^i (\xx)}
    \frac{\delta \fibCi^i(\xx)}{\delta y^k(\XX)} \,
  \d \xx
  +
  \dots
\end{align}
Here we intentionally dropped the derivatives of $F$
with respect to the fields $(\rho,\mm,\AAA$).
Putting \eqref{eq:DerrGM}, \eqref{eq:DerrFy} and \eqref{eq:DerrFibCy} together we obtain\footnote
{
	In the sense of \cite{Morrison1998}, the integration represents a~duality,
	the derivative of one field with respect to other an operator from one tangent space to another,
	and the Fubini theorem to an adjoint operation in the scalar product.
}
\begin{align*}
  &\int_{\RRR^3}
    \frac{\delta B}{\delta y^k(\XX)}
    \frac{\delta C}{\delta M_k(\XX)} \,
  \d \XX \\
  &=
    \int_{\RRR^3} \int_{\RRR^3}\left[
   \frac{\delta B}{\delta \fibCi^i(\xx)}
   \left(
     - \frac{\partial \delta}
           {\partial X^I}(\XX-\yy^{-1}(\xx))
     \delta^i_k \fibRi^I(\xx)
     - \frac{\partial \fibCi^i}{\partial x^k} (\xx) \delta(\XX-\yy^{-1}(\xx))
   \right)
   \left.
    \frac{\delta C}{\delta m_k(\xx)}\right]
  \right|_{\XX = \yy^{-1}(\xx)} \,
  \d \xx \,
  \d \XX \\
   &=
   -
   \int_{\RRR^3}
     \frac{\delta B}{\delta \fibCi^i(\xx)}
     \fibRi^I(\xx)
     \int_{\RRR^3}
       \frac{\partial \delta}
           {\partial X^I}(\XX-\yy^{-1}(\xx))
     \left.
       \frac{\delta C}{\delta m_i(\xx)}
     \right|_{\XX = \yy^{-1}(\xx)} \,
     \d \XX \,
   \d \xx
   -
   \int_{\RRR^3}
     \frac{\partial \fibCi^i}{\partial x^k} (\xx)
     \frac{\delta B}{\delta \fibCi^i(\xx)}
     \frac{\delta C}{\delta m_k(\xx)} \,
    \d \xx \\
  &=
  \int_{\RRR^3}
     \frac{\delta B}{\delta \fibCi^i(\xx)}
       \fibRi^I(\xx)
       \frac{\partial}{\partial X^I}
       \frac{\delta C}{\delta m_i(\xx)} \,
    \d \xx
   -
   \int_{\RRR^3}
     \frac{\partial \fibCi^i}{\partial x^k}(\xx)
     \frac{\delta B}{\delta \fibCi^i(\xx)}
     \frac{\delta C}{\delta m_k(\xx)} \,
    \d \xx \\
  &=
  \int_{\RRR^3}
     \frac{\delta B}{\delta \fibCi^i(\xx)}
       \frac{\partial x^j}{\partial X^I}
       \fibRi^I(\xx)
       \frac{\partial}{\partial x^j}
       \frac{\delta C}{\delta m_i(\xx)} \,
    \d \xx
   -
   \int_{\RRR^3}
     \frac{\partial \fibCi^i}{\partial x^k}(\xx)
     \frac{\delta B}{\delta \fibCi^i(\xx)}
     \frac{\delta C}{\delta m_k(\xx)} \,
    \d \xx \\
  & \quad
  =
  \int_{\RRR^3}
    \fibCi^j(\xx)
    \frac{\delta B}{\delta \fibCi^i(\xx)}
    \frac{\partial}{\partial x^j}
    \frac{\delta C}{\delta m_i(\xx)} \,
  \d \xx
   -
   \int_{\RRR^3}
     \frac{\partial \fibCi^i}{\partial x^k}(\xx)
     \frac{\delta B}{\delta \fibCi^i(\xx)}
     \frac{\delta C}{\delta m_k(\xx)} \,
    \d \xx.
\end{align*}
Since the right-hand side depends solely on~the new state variables,
the projection was successful.
By renaming the indices, we can now write down bracket \eqref{eq.PB.Bb}.

\end{document}